\documentclass{emulateapj}

\newcommand\cvir{{\hbox{$c_{\rm vir}$}}}
\newcommand\rvir{{\hbox{$r_{\rm vir}$}}}
\newcommand\mvir{{\hbox{$M_{\rm vir}$}}}
\newcommand\msun{\hbox{{$M_{\odot}$}}}
\newcommand\chandra{{\sl Chandra}}
\newcommand\xmm{{\sl XMM}}
\newcommand\rosat{{\sl ROSAT}}
\newcommand\einstein{{\sl Einstein}}
\newcommand\omegam{\hbox{{$\Omega_{\rm m}$}}}
\newcommand\omegalambda{\hbox{{$\Omega_{\Lambda}$}}}
\newcommand\apec{{\sc apec}}
\newcommand\fe{\hbox{{$Z_{\rm Fe}$}}}

\shorttitle{DM radial profile of A2589}
\shortauthors{Zappacosta et al.}

\begin{document}

\title{The Absence of Adiabatic Contraction of the Radial Dark Matter
\\ Profile in the Galaxy Cluster A2589} 

\author{Luca Zappacosta\altaffilmark{1}, David
  A. Buote\altaffilmark{1}, Fabio Gastaldello\altaffilmark{1}, Phillip
  J. Humphrey\altaffilmark{1}, James Bullock\altaffilmark{1}, Fabrizio
  Brighenti\altaffilmark{2,3} and William Mathews\altaffilmark{2}}
\altaffiltext{1}{Department of Physics and Astronomy, University of California,
    Irvine, 4129 Frederick Reines Hall, Irvine, CA 92697-4575;\\
    lzappaco@uci.edu, buote@uci.edu}
\altaffiltext{2}{UCO/Lick Observatory, Board of Studies in Astronomy and
  Astrophysics, University of California, Santa Cruz}
\altaffiltext{3}{Dipartimento di Astronomia, Universit\`a di Bologna,
  Bologna, Italy}

\begin{abstract}
We present an X-ray analysis of the radial mass profile of the
radio-quiet galaxy cluster A2589 between $0.015-0.25\,\rm{r_{vir}}$
using an XMM-Newton observation. Except for a $\approx16$~kpc shift
of the X-ray center of the $R=45-60$~kpc annulus, A2589 possesses a
remarkably symmetrical X-ray image and is therefore an exceptional
candidate for precision studies of its mass profile by applying
hydrostatic equilibrium. The total gravitating matter profile is well
described by the NFW model (fractional residuals $\lesssim10\%$) with
$c_{\rm vir}=6.1\pm0.3$ and $M_{\rm vir}=3.3\pm0.3\times10^{14}M_{\odot}$ ($r_{vir}=1.74\pm0.05$~Mpc) in excellent
agreement with $\Lambda$CDM. When the mass of the hot ICM is
subtracted from the gravitating matter profile, the NFW model fitted
to the resulting dark matter (DM) profile produces essentially the
same result. However, if a component accounting for the stellar mass
($M_*$) of the cD galaxy is included, then the NFW fit to the DM
profile is substantially degraded in the central $r\sim50$~kpc for
reasonable $M_*/L_V$. Modifying the NFW DM halo by adiabatic
contraction arising from the early condensation of stellar baryons in
the cD galaxy further degrades the fit. The fit is improved
substantially with a Sersic-like model recently suggested by high
resolution N-body simulations but with an inverse Sersic index,
$\alpha\sim0.5$, a factor of $\sim3$ higher than predicted.  We argue
that neither random turbulent motions nor magnetic fields can provide
sufficient non-thermal pressure support to reconcile the \xmm\ mass
profile with adiabatic contraction of a CDM halo assuming reasonable
$M_*/L_V$. Our results support the scenario where, at least for
galaxy clusters, processes during halo formation counteract adiabatic
contraction so that the total gravitating mass in the core
approximately follows the NFW profile.
\end{abstract}


\section{Introduction}

The properties of dark matter (DM) halos are a powerful discriminator
between different cosmological models.  Of particular importance is
the distribution of halo concentration ($\cvir$) with virial mass
($\mvir$).  For the concordance cold dark matter ($\Lambda$CDM)
cosmology the mean $\cvir$ varies slowly over a factor of 100 in
$\mvir$, whereas the scatter remains very nearly constant
\citep[e.g.,][]{bull01a,kuhl05a}. At fixed halo mass, the distribution
of concentrations is expected to vary significantly as a function of
the cosmological parameters, including $\sigma_8$, $n$, and $w$, the
dark energy equation of state \citep[e.g.][]{dola04a,kuhl05a}.

The radial density profiles of CDM halos are fairly well described
between approximately $0.01-1\,\rvir$ (where $\rvir$ is the virial
radius) by the 2-parameter ``NFW'' model suggested by
\citet{nfw}. More recent numerical simulations with higher resolution
show that CDM halos deviate slightly from this average NFW profile
because the density slope changes continuously with radius
\citep[e.g.,][]{nava04a,diem04a,grah05a}. However, there is not yet a
general consensus on whether the central density slope is shallower or
steeper than the NFW profile.
                                                                                
The central regions of DM halos also reveal vital information about
the interaction between the stellar baryons and the DM during halo
formation. If the stellar baryons condense much earlier than the
dissipationless DM, it is expected that the baryons will adiabatically
compress the DM halo away from its pure NFW form
\citep[e.g.,][]{blum86a,gned04a}. It has also been argued that heating
of the dark matter by dynamical friction with member galaxies
counteracts adiabatic compression and leads to a total gravitating
mass profile consistent with the pure NFW profile
(e.g., \citealt{elza04a}; see also \citealt{loeb03a}). 

In theory, galaxy clusters are excellent sites to study DM profiles,
particularly in their cores, because they are DM-dominated deep down
to a small fraction of the virial radius\citep[$\approx 0.007\rvir$;
e.g.,][]{dubi98a,lewi03a} and several powerful techniques exist to
probe cluster mass profiles. In practice, however, it has proven quite
difficult to obtain precise, reliable measurements of the DM profiles
in cluster cores. Stellar dynamical studies suffer from the
long-standing and well-known problem of velocity dispersion
anisotropy, and despite occasional claims to the contrary
\citep[e.g.,][]{sandd04}, cannot obtain precise constraints without
restrictive assumptions on the form of the velocity dispersion tensor,
and therefore of the mass profile. Since the weak lensing
approximation breaks down within the central $\approx 100$~kpc of
clusters, only those clusters with giant arcs produced by strong
lensing can be used to map the core mass distribution. Unfortunately,
giant arcs are preferentially produced in clusters with significant
substructure in the core, making them much less suitable for
comparison to the average relaxed cluster profile predicted by
simulations. 

X-ray observations of the hot intracluster medium (ICM) are a powerful
probe of the DM in galaxy clusters \citep[e.g., see recent reviews
by][]{buot04d,arna05a}. Since the pressure tensor of the hot ICM is
isotropic, and the gas traces the three-dimensional cluster potential
well within the virial radius, X-ray observations are a vital
complement to gravitational lensing techniques that probe only the
projected potential. However, the X-ray method requires the ICM to be
in approximate hydrostatic equilibrium. Morphological studies of X-ray
images have shown that a large fraction of nearby clusters ($z<0.2$)
do have regular image morphologies and appear to be nearly relaxed on
0.5-1 Mpc scales \citep[e.g.,][]{mohr95a,buot96b,jone99a}, though no
cluster is observed to be (or is expected to be) devoid of evidence of
disturbance.

Precisely how departures from hydrostatic equilibrium translate to
errors in mass estimates is as yet not well understood in terms of
quantifiable measures of the X-ray image morphology, projected
temperature, or projected metallicity maps. What is clear is that
errors in mass estimates should be minimized for clusters without
obvious substructure \citep[][though see
\citealt{hallman}]{tsai,buot95a,evrard}. That is, clusters with
regular X-ray isophotes, centrally peaked cores and no central
AGN-induced disturbance. Further evidence for a relaxed hot ICM is
suggested by a temperature profile that rises with radius from a
minimum in the cluster core \citep[e.g.,][]{degr02a} and by
metallicity profile that is peaked at the center and declines
monotonically with increasing radius
\citep[e.g.,][]{degr04a,boehringer}.

Unfortunately, it has proven difficult to find cluster candidates that
are relaxed both on large ($\approx 0.5-1$~Mpc) scales and within
their cores because evolved, cool-core clusters typically have X-ray
and radio disturbances within their cores believed to arise from AGN
feedback \citep[e.g.,][]{birz04a}. A small number of radio-quiet
clusters, previously classified as cooling flows, have been observed
with \chandra; A2589 \citep{buot04a}, A644 \citep{buot05a}, A1650 and
A2244 \citep{dona05a}. None of these clusters displays the central
minimum in the temperature profile characteristic of cool-core
clusters or has any evidence for AGN feedback. The lack of AGN
evidence for current feedback in these systems may arise from heating
due to merging \citep{buot05a} or unusually powerful AGN feedback
events in the distant past \citep{dona05a}.

The cluster A2589 ($z=0.0414$) is observed to be one of the most
morphologically regular clusters in terms of its X-ray emission on a
scale of $\approx 0.5$~Mpc \citep{buot96b}. The high-resolution
\chandra\ image confirmed a highly symmetrical X-ray image morphology
down to the center, although with some evidence for a small isophotal
center shift of $\sim 10$~kpc. The highly regular X-ray image
morphology combined with no evidence for AGN feedback make this
cluster an excellent candidate for studies of its mass profile by
making the approximation that the hot gas is in hydrostatic
equilibrium. Because of the low photon statistics of our shallow
\chandra\ observation we were unable to place strong constraints on
the mass profile, though an NFW profile was found to be consistent
with the data within $r\sim 150$~kpc.

In this paper we improve upon our previous study by using a higher
quality XMM-Newton observation to extend our analysis to
$\sim400\,\rm{kpc}$ corresponding roughly to $\sim
1/4\,\rm{r_{vir}}$. The higher quality data will allow us to test
different mass profiles proposed so far in the literature to account
for the dark matter and the influence of the central bright dominant
galaxy.

Our assumed virial radius is defined as the radius of a sphere whose
mean density is 104.7 times the critical density of the universe. This
value is estimated at the redshift of the cluster and for an
$(\Omega_m,\Omega_{\Lambda})=(0.3,0.7) $ universe \citep{bryan}.  (The
redshift of A2589 corresponds to an angular diameter distance of
171~Mpc and $1\arcsec = 0.82$~kpc assuming $\omegam=0.3$ and
$\omegalambda=0.7$.)

\section{Observation and Data Preparation}
\label{obs}

A2589 was observed by XMM-Newton for $\sim 46$~ks during revolution
822. The observation was performed in full window mode with the three
EPIC cameras equipped with the thin filter. For the data reduction we
used SAS~6.0, and for the spectral analysis we used XSPEC~11.3.1.

The observation suffered from several periods of background flaring.
First, we excluded point sources that were identified by visual
inspection. Then we isolated affected time intervals from visual
inspection of light curves extracted in both high-energy (10-12~keV
for the EPIC-MOS and 10-13~keV for the EPIC-PN) and low-energy
(0.5-2.0~keV) bands.  The light curves were extracted from regions far
from the center of A2589.  After excising the flaring time intervals
we arrived at cleaned exposures times of $\sim16.6\,\rm{ks}$ (MOS1),
$\sim17.6\,\rm{ks}$ (MOS2), and $\sim12.8\,\rm{ks}$ (PN). (Note only
single events were used for the PN.)

Since the cluster emission covers all the EPIC field of view it is not
possible to obtain a local background estimate without accounting for
source contamination. The standard procedure in this case is to use
background templates obtained from nominally blank fields
\citep{read}. But these standard templates are not appropriate in regions at large
radii where the source emission is comparable to the background
because of CXB spatial variations and, possibly, residual flaring
particular to a given exposure. Consequently, we estimate the
background by simultaneously fitting a model consisting of source and
background components to spectra as far away from the source center as
possible.

Our procedure to subtract the background is an small update of the
approach summarized in \citet{buot04b} that is explained fully in
\citet{buote06,gastaldello}. In sum, the model consists of components
for (1) the cluster emission, (2) the Cosmic X-ray Background (CXB)
represented by two components for the soft thermal foreground emission
and one power-law model for the extragalactic non-thermal background,
and (3) the quiescent instrumental background consisting of a
continuous spectral component represented by a broken power-law (whose
parameters were determined from fitting the out-of-field of view
events) and several gaussians to account for the instrumental
fluorescent lines.  We model the non-thermal extragalactic CXB
component due to unresolved AGNs using a power-law with a slope
$\Gamma=1.41$. For galaxy clusters, because of their high gas
temperatures, the contribution of the hot gas emission is partially
degenerate with this high-energy power law component. Therefore, we
required that the normalization of the CXB power-law component was
initially set to that obtained by
\citet{deluca} and left free to vary only within the cosmic variance
\citep{barcons}.  All the cosmic components are absorbed by the
Galactic column density value measured in the cluster region
$\rm{N_{H}}=4.15\times10^{20}\,\rm{cm^{-2}}$
\citep{dick90}. Finally, we noticed that the broken power-law that we use to
approximate the instrumental background of the PN detector does not
fit well the high-energy portions of the spectra in all of the annuli
simultaneously. (See \S \ref{spec_analysis} for annuli definitions.)
This may be ascribed to a residual flaring component of soft protons
not completely removed that appears only in the inner regions as it is
focused and vignetted by the satellite optics
\citep{read}. We were able to obtain an acceptable fit if
we allowed the shape of the broken power-law in the outermost annulus 
to vary separately from the other annuli.

Response matrices were generated for each region using standard SAS
tasks, which enable the generation of photon-weighted ancillary
response (ARF) files. However, these tools only enable redistribution
matrix files (RMFs) to be generated appropriate for individual chips
of the PN. Since we desire to analyze circular annuli that, in
general, cover multiple chips, we created a photon-weighted RMF for
each annulus combining the RMFs generated for appropriate ranges of
CCD rows, between which the response is known to vary
appreciably. This weighting was not performed for the MOS since the
SAS-produced MOS RMFs do not vary over the field of view.
\begin{figure*}[t]
\parbox{0.49\textwidth}{
\centerline{\includegraphics[angle=0,width=0.45\textwidth]{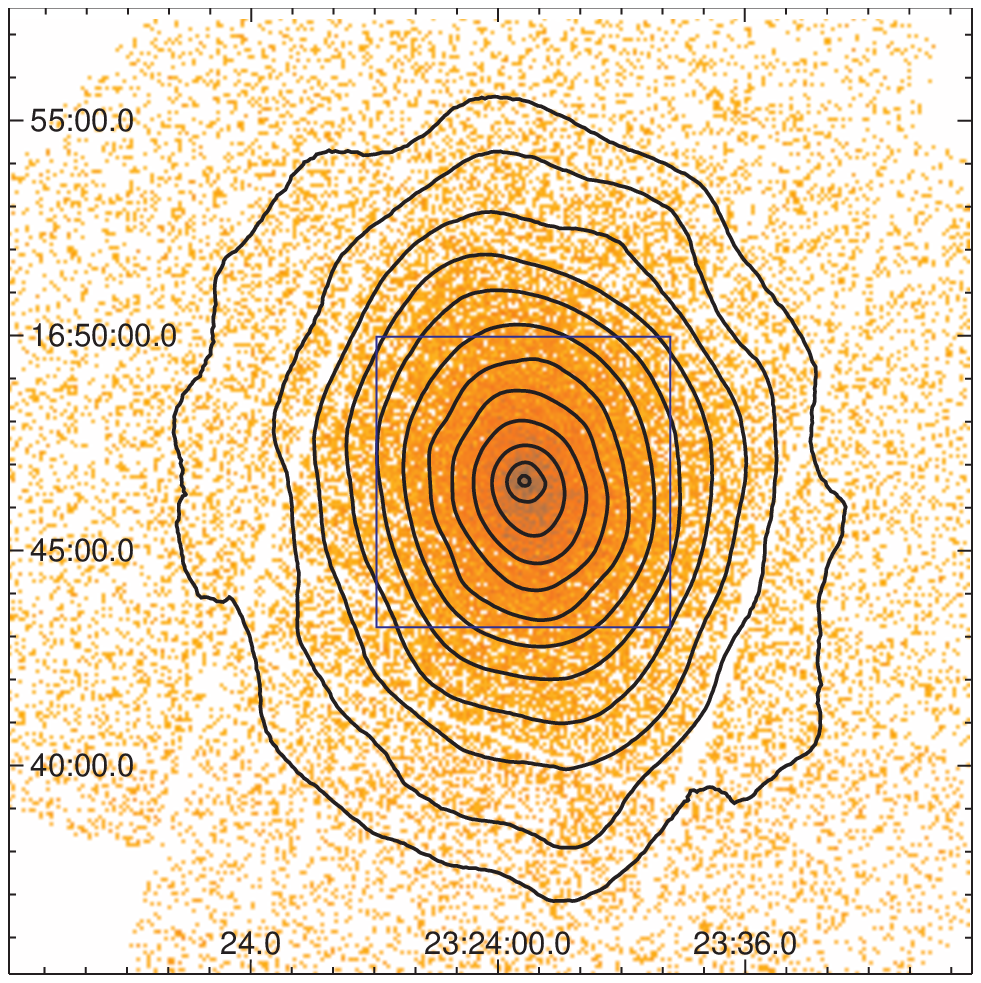}}}
\parbox{0.49\textwidth}{
\centerline{\includegraphics[angle=0,width=0.45\textwidth]{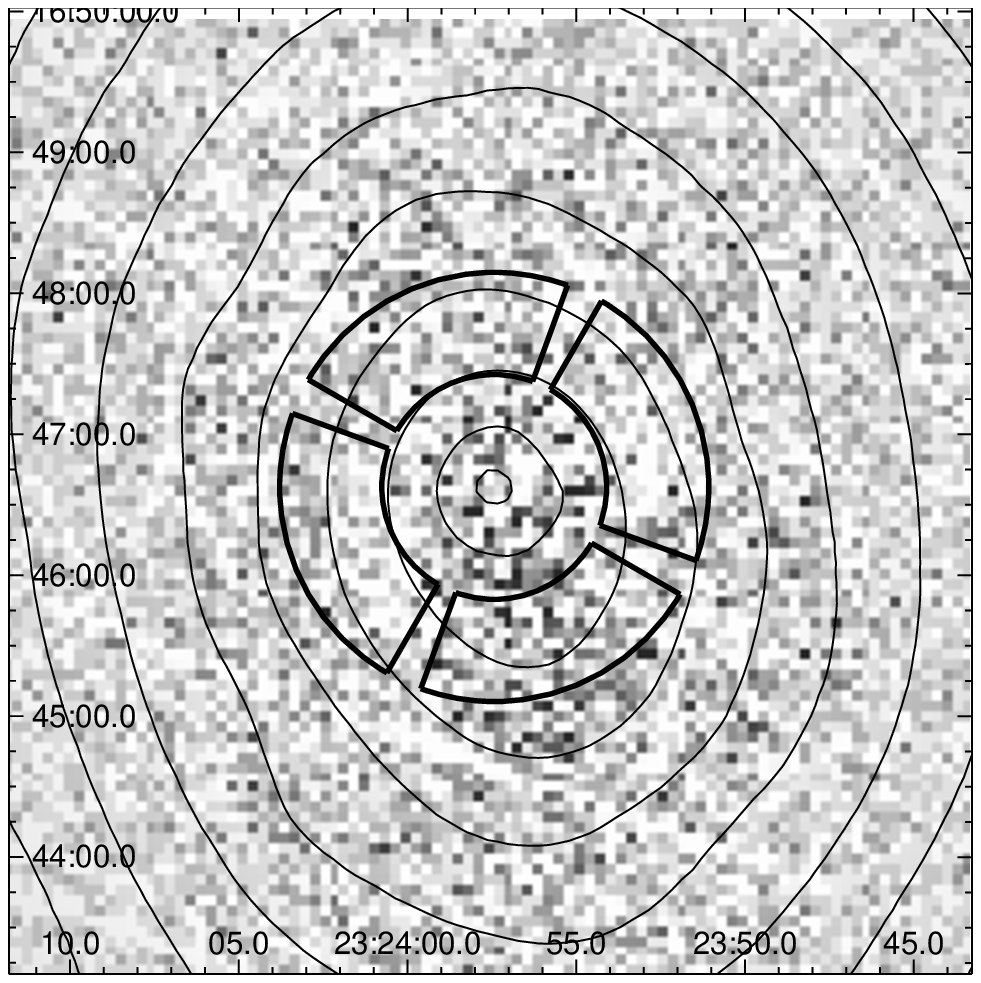}}}
\caption{\label{cluster_images} \footnotesize
({\it Left}) MOS1 image in the 0.5-8.0~keV band displayed with
logarithmic intensity scaling. This image has not been exposure
corrected due to the inaccurate exposure maps available from the
standard processing. No additional smoothing has been applied to this
image. However, smoothed intensity contours are overlaid to guide the
eye. ({\it Right}) Residual image of MOS1, displayed with linear
intensity scaling, obtained by subtracting from the raw image a model
constructed by fitting perfect elliptical isophotes (see text). The
region displayed corresponds to the boxed area of the image on the
left. The sectors indicate regions analyzed to assess non-ellipsoidal
fluctuations as discussed in the text.}
\end{figure*}

\begin{figure}[t]
  \begin{center} \includegraphics[width=0.45\textwidth,
  angle=0]{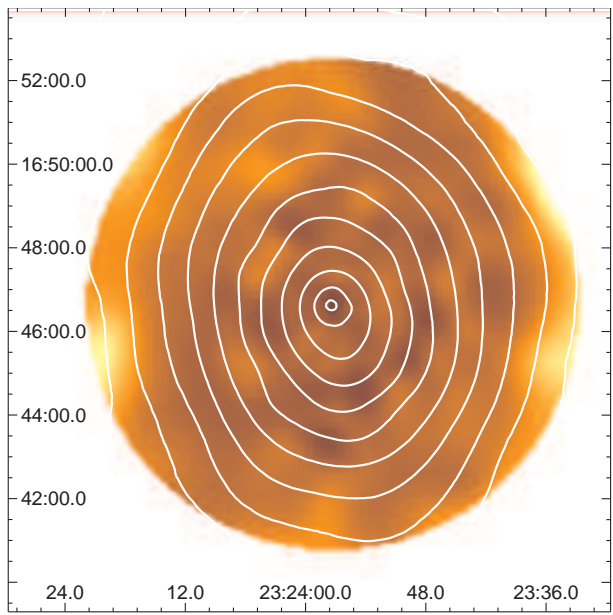} \caption{Hardness ratio image of A2589,
  displayed with logarithmic intensity scaling, generated using the
  0.5-1.3~keV and 1.3-5.0~keV combined MOS1 and MOS2 images (see text
  for details). Contours represent the X-ray isophotes.}  \label{hr}
  \end{center}
\end{figure}

\section{Imaging analysis}\label{image_analysis}

We display the MOS1 image in Figure~\ref{cluster_images}. The image is
not exposure corrected because the standard processing does not
provided accurate exposure maps. Since the MOS2 and pn images have
more chip gaps and bad columns, we only display the MOS1 image.  The
diffuse emission is remarkably regularly shaped, being moderately
elliptical, and fills the entire field of view. We measure the
centroid and the ellipticity of the X-ray surface brightness using the
moment method described by \citet{cm} and implemented in our previous
X-ray studies of galaxies and clusters \citep[e.g.,][]{buot94}. This
iterative method is equivalent to computing the (two-dimensional)
principal moments of inertia within an elliptical region. The
ellipticity is defined by the square root of the ratio of the
principal moments, and the position angle is defined by the
orientation of the larger principal moment. Following our previous
study of the ellipticity of the \chandra\ data of NGC 720
\citep{buot02b} we removed point sources and replaced them with
smoothly distributed diffuse emission using the {\sc CIAO} task {\sc
dmfilth}. Because this method for computing ellipticity cannot account
for chip artifacts we restricted the analysis to the central chip on
the MOS1.

We defined the following annuli with outer semi-major axes $(a)$
expressed in units of kpc (15, 30, 45, 60, 75, 100, 150, 225). The
ellipticity is rather uncertain within $a\sim 30$~kpc because of the
small number of pixels (and resolution elements). For $a\ga 60$~kpc
the ellipticity falls from near 0.30 to 0.25 at the outer radius with
a position angle near $10\degr$ N-E over this entire range.

The intensity weighted center shift, defined similarly to that in
\citet{mohr95a}, is $5.4\pm 0.5$~kpc for these annuli. However, the
maximum difference between annuli centers is observed between the
annulus 45-60~kpc and the central circle (0-15~kpc) where we obtain a
difference of 16~kpc in center positions with negligible statistical
error. This 45-60~kpc annulus appears to be slightly offset from all
of the other annuli. Note that the X-ray peak (R.A.: 23h23h57.4m,
DEC: 16d46m38s) coincides precisely with the optical position of the
cD galaxy NGC 7647.

The most general non-rotating, self-gravitating equilibrium
configuration is the triaxial ellipsoid. The equipotentials of a
triaxial ellipsoid are concentric surfaces that themselves are nearly
ellipsoids. Since in hydrostatic equilibrium the hot gas traces
exactly the same shape as the gravitational potential regardless of
the radial temperature profile \citep{buot94,buot96a}, we expect a
relaxed cluster to have concentric isophotes that are nearly
elliptical in shape, though the ellipticity may vary with radius. (The
position angle of the X-ray isophotes also need not be constant with
radius.)

The small center shift we measured above represents a deviation from
equilibrium, particularly near the isophotes with $a=45-60$~kpc. We
searched for higher order deviations from elliptical symmetry in the
following manner. Using the {\it ellipse} package within {\sc
iraf-stsdas} we fitted elliptical isophotes to the MOS1 image; we
restricted this analysis to the central CCD to avoid complications
associated with the chip gaps. Then we constructed a model image from
the fitted isophotes and subtracted it from the raw image. The
resulting residual image is also displayed in Figure
~\ref{cluster_images}. We compared the model image and raw image in 4
sectors defined near the 45-60~kpc annulus; i.e., inner radius 40~kpc,
outer radius 75~kpc. For each sector we obtain the following ratios of
counts between the raw and model images: North $1.01\pm0.02$, South
$1.04\pm0.02$, East $0.92\pm0.02$, West $0.94\pm0.02$. These values
represent quite reasonable scatter about unity considering uncorrected
exposure variations. We conclude the image does not show significant
deviations from elliptical symmetry apart from the modest center shift
noted above.

We also searched for evidence of asymmetry in the temperature (and
metallicity) structure using a hardness ratio map. Using the combined
MOS1 and MOS2 images we constructed the map as an (S-H)/(S+H) image,
where S and H are respectively images in soft (0.5-1.3~keV) and hard
(1.3-8.0~keV) bands; the image was then smoothed with a gaussian
kernel with width of 1 pixel. We display the hardness-ratio map in
Fig.~\ref{hr}. It is quite regular and shows no evidence for
significant deviations from elliptical symmetry.

As first reported by \citet{davi96a} there is a strong alignment
between the optical isophotes of the cD, the X-ray isophotes, and the
galaxy isopleths of the host (super-) cluster. It is also worth noting
that the average ellipticity of the cluster galaxy isopleths is $\sim
0.3$ \citep{plio91a}, consistent with the X-ray isophotes. \citet{beer91a}
reported an offset of the cD velocity ($\sim 250$~km~s$^{-1}$) from the
other member galaxies. However, these sparse data were deemed
insufficient to obtain a robust result, and, consequently, this system
was not included in the systematic study of cD offsets by
\citet{bird94a}. Indeed, the kinematical study by \citet{gebh91a}
using the data of \citet{beer91a} obtains only weak evidence for
substructure. 

Evidently, A2589 is nearly in hydrostatic equilibrium. This is
indicated by the strong similarity in the distributions of galaxies
and ICM, and because A2589 has one of the most regular X-ray images on
the $\approx 0.5$~Mpc scale \citep{buot96b}.  The only significant
asymmetry we detect is the small center offset associated with the
45-60~kpc annulus mentioned above. We will consider whether this
region displays anomalous behavior in our analysis of the azimuthally
averaged spectral (and mass) properties below.

\renewcommand{\arraystretch}{1.25}
\setlength{\tabcolsep}{2.5pt}
\begin{deluxetable*}{@{\extracolsep{\fill}}cccccccccc}
\tablecaption{Spectral fit parameters for each annulus \label{annuli}}
\tabletypesize{\scriptsize}
\tablehead{
      \colhead{Annulus}  &  \colhead{$\rm{R_{in}}$}     &  \colhead{$\rm{R_{out}}$}    &   \colhead{T}     &  \colhead{$\rm{Z_{Fe}}$}  &  \colhead{$\rm{Z_{S}}$}   &  \colhead{$\rm{Z_{Si}}$}     &  \colhead{norm}                     &  \colhead{$\rm{\chi^{2}/dof}$}    &  Null Hyp.                \\
      \colhead{}         &  \colhead{(kpc/arcmin)}      & \colhead{(kpc/arcmin)}       &   \colhead{(keV)} &  \colhead{(solar)}        &  \colhead{(solar)}        &  \colhead{(solar)}           &  \colhead{($10^{-3}\rm{cm^{-5}}$)}  &                                   &  prob. (\%)
}
\startdata                                                           
1 &\,\,0/0.00 & \,32/0.65 & $3.26 \pm 0.08$ & $1.33 \pm 0.11$ & $0.55 \pm 0.25$ & $1.01 \pm 0.20$ & $1.38 \pm 0.04$ & 535.3/487 &  6.39\\
2 & \,32/0.65 & \,57/1.16 & $3.45 \pm 0.08$ & $0.95 \pm 0.07$ & $0.17 \pm 0.16$ & $0.47 \pm 0.19$ & $2.10 \pm 0.04$ & 654.6/606 &  8.38\\
3 & \,57/1.16 & \,93/1.89 & $3.61 \pm 0.07$ & $0.78 \pm 0.06$ & $< 0.22       $ & $0.30 \pm 0.16$ & $3.42 \pm 0.08$ & 752.0/730 &  27.8\\
4 & \,93/1.89 & 136/2.76 & $3.40 \pm 0.05$ & $0.77 \pm 0.06$ & $0.14 \pm 0.09$ & $0.16 \pm 0.08$ & $3.72 \pm 0.07$ & 772.9/716  &  6.91\\
5 & 136/2.76 & 193/3.92 & $3.36 \pm 0.06$ & $0.54 \pm 0.05$ & $0.22 \pm 0.12$ & $0.52 \pm 0.09$ & $3.99 \pm 0.06$ & 653.6/688   &  82.0\\
6 & 193/3.92 & 289/5.87 & $3.52 \pm 0.07$ & $0.57 \pm 0.07$ & $< 0.12       $ & $< 0.14       $ & $4.56 \pm 0.09$ & 726.4/676   &  8.53\\
7 & 289/5.87 & 489/9.93 & $3.04 \pm 0.15$ & $0.43 \pm 0.07$ & $0.30 \pm 0.15$ & $0.28 \pm 0.19$ & $5.40 \pm 0.16$ & 493.5/450   &  7.12
\enddata
\tablecomments{Parameters derived by fitting an absorbed {\sc apec}
plasma model to the spectra extracted in each annulus.  Columns
$Z_{Fe}$, $Z_{S}$, $Z_{Si}$ show the abundance values obtained leaving
free to vary Fe, S and Si. The ``norm'' parameter is the emission
measure for the {\sc apec} code as defined in XSPEC.  The last column
report the Null Hypothesis Probability for the fit. The total
0.5-8.0~keV luminosity of these annuli is $1.47^{+0.18}_{-0.16}\,
10^{43}$~erg~s$^{-1}$ (90\% conf.).}
\end{deluxetable*}

\section{Spectral analysis}\label{spec_analysis}

Since the X-ray image and hardness-ratio map do not display
substantial azimuthal variations, we focus our analysis on the X-ray
spectral properties derived from circular annuli. We neglect the
ellipticity of the X-ray isophotes since determining the flattening of
the mass distribution is beyond the scope of this paper. (In any
event, using elliptical annuli does not produce qualitatively
different profiles of the gas density, temperature, or metallicity when
the results are expressed in terms of the mean radius $R=a\sqrt{q}$,
where $a$ is the semi-major axis and $q$ is the axial ratio.)

We extracted spectra in the 0.5-8.0~keV energy band from all of the
EPIC CCDs in a series of concentric circular annuli centered on the
X-ray peak. These annuli were originally constructed to have at least
8000 background-subtracted counts in the MOS1 and were defined to be
larger than the \xmm\ PSF and provide temperature constraints of
similar precision.  Since the central radial bin has the highest S/N,
we further divided it into two annuli (each still being larger than
the \xmm\ PSF). This allows us to probe the spectral properties down
to $\sim 0.015\rvir$. The annuli definitions are listed in Table
\ref{annuli}.  We do not list results obtained for the outermost annulus
($R>9.93\arcmin$ out to the edge of the fields) that were used in the
background determination summarized in \S
\ref{obs}. At these large radii the source emission contributes little
to the EPIC spectra, and thus small changes in our adopted background
model translate to relatively large changes in the derived source
parameters. (We assess the systematic error resulting from the
background level on our mass results in \S \ref{systematics_sect}.)
We also restricted the upper energy limit to 5.0~keV in the
penultimate annulus (i.e., the last annulus listed in Table
\ref{annuli}) because of excess hard emission that we could not remove
with our background model.

In each annulus we fitted a model consisting of an optically thin hot
plasma (\apec) modified by Galactic absorption. The free parameters
are the normalization, temperature, iron, silicon, and sulfur
abundances. All other elemental abundances are fixed so that their
ratio with respect to iron is solar assuming the abundance standard of
\citet{grevesse}. Error estimates are obtained by simulating spectra
based on our best-fitting models. From 20 Monte Carlo simulations we
compute the standard deviation of each parameter which we report as
the $1\sigma$ error. (All quoted errors are $1\sigma$ unless stated
otherwise.)

\begin{figure*}[t]
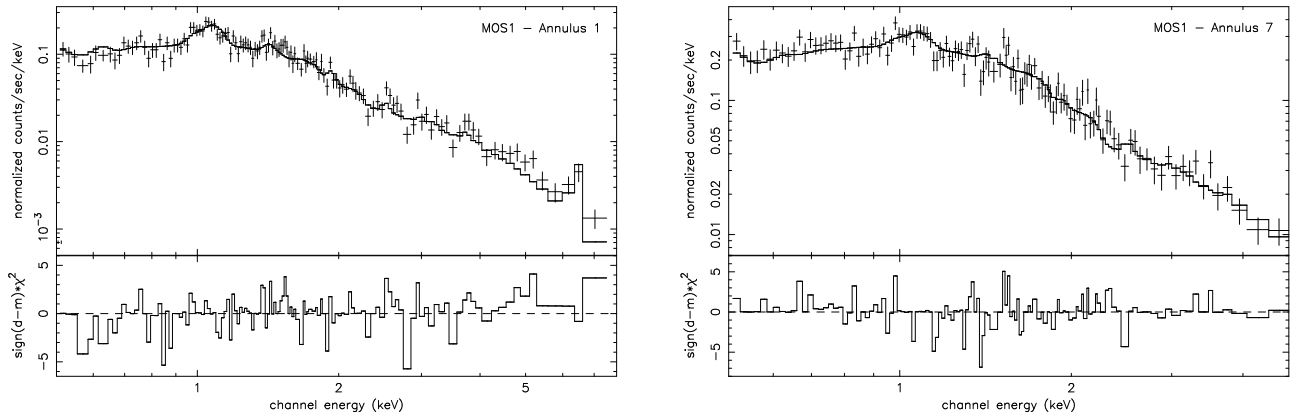

\parbox{0.49\textwidth}{
\centerline{\includegraphics[angle=-90, width=.45\textwidth]{f3a.ps}}
}
\parbox{0.49\textwidth}{
\centerline{\includegraphics[angle=-90, width=.45\textwidth]{f3b.ps}}
}
\caption{\label{fig.spec} MOS1 spectra accumulated within
({\sl Left Panel}) annulus \#1 and ({\sl Right Panel}) annulus
\#7. Each spectrum is fitted with an \apec\ plasma model modified by
Galactic absorption as discussed in \S \ref{spec_analysis}. Note that
these models were fitted simultaneously with the MOS2 and pn data,
but we display only the MOS1 for clarity.}
\end{figure*}

The results are listed in Table \ref{annuli} for the model parameters,
and we display the MOS1 spectra and best-fitting models for two annuli
in Figure \ref{fig.spec}.  The quantities we derive for the hot gas in
each annulus are average values weighted by cluster emission projected
along the line-of-sight.  Below we will obtain the gas density and
temperature as a function of three-dimensional radius by projecting
models along the line-of-sight and fitting them to the results
obtained as a function of projected radius listed in Table
\ref{annuli}. We found this procedure to be appropriate given the
small number of radial bins and quality of our data.  In \S
\ref{systematics_sect} we also consider the effects on the derived
mass profile of first deprojecting the data using the well-known
onion-peeling method.

We mention that the generally sub-solar ratios of Si/Fe and S/Fe
abundances imply that most of the iron ($\sim 80\% - 95\%$) in the hot
ICM has been provided by Type~Ia supernova, with convective
deflagration explosion models favored over delayed-detonation
models. These results are consistent with those we have obtained for
galaxies and groups \citep[e.g.,][]{buot03b,humphrey06a}.

\begin{figure}[t]
  \begin{center} \includegraphics[width=0.5\textwidth,
  angle=0]{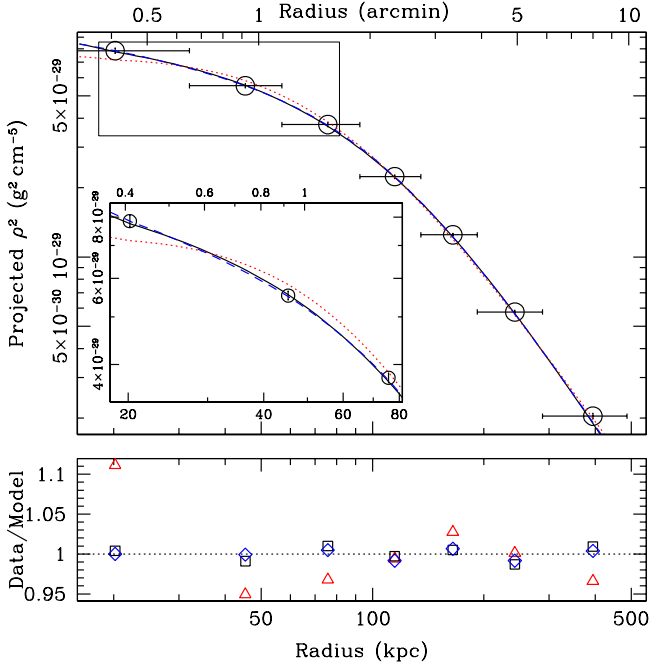} \caption{The radial profile of
  the projection of $\rho_g^2$. The data points are proportional to
  $norm/A$, where $norm$ is listed in Table \ref{annuli} and $A$ is
  the area of the annulus. The models have been fitted by weighting
  the projection integral by the plasma emissivity (see \S
  \ref{density}). However, for display purposes we do not bin the
  models in the plot.  Because the statistical error bars are very
  small, we enclose the data points with open circles. Each data point
  represents a radial bin as defined in Table \ref{annuli}. The
  horizontal bars represent the width of the annuli. The solid line is
  the best fit cusp~$\beta$~model, the dotted red line is the single
  $\beta$~model and the dashed blue line is the
  double~$\beta$~model. The cusp~$\beta$~model and
  double~$\beta$~model almost entirely overlap. The inset shows an
  enlarged view of the inner region. In the ratio (data/model) plot,
  red triangles, blue diamonds and black squares refer respectively to
  $\beta$~model, double $\beta$~model and cusped $\beta$~model.}
  \label{density_profile}
\end{center}
\end{figure}

\section{Gas Density Profile}
\label{density}

\begin{deluxetable*}{@{\extracolsep{\fill}}lll|ll}
    \tablecaption{Gas Density and Temperature Models\label{proj_prof}}
\tabletypesize{\scriptsize}
\tablehead{
\multicolumn{3}{c|}{Projected Density} & \multicolumn{2}{c}{Projected Temperature}     \\
  $\beta$~model                & double~$\beta$~model                  & \multicolumn{1}{l|}{cusp~$\beta$~model}                &     Power-law~model                        & $T(r)^*$~model                      
}
\startdata
  $\beta\!=\!0.54\pm0.01$      & $\beta\!=0.57\pm0.02$  & $\beta\!=\!0.59\pm0.02$    &     $T_{100}\!=\! 3.42\pm0.03\,\rm{keV}$   & $A_{1}\!=\! 3.6\pm1.5$              \\
  $r_c\!=\!76.5\pm2.4$~kpc     & $r_{c1}\!=\!21.9\pm14.5$~kpc          & $r_{c}\!=\!119.2\pm12.5$~kpc      &     $\alpha_p \!=\!  0.0\pm0.02$           & $A_{2}\!=\! 3.5\pm0.9$              \\
  $\rho_{g0}=1.40\pm 0.03^{\ast}$                      & $r_{c2}\!=\!95.2\pm14.0$~kpc           & $\alpha_c\!=\!0.38\pm0.06$       &     \nodata                                & $r_p=100$(fixed)                    \\
  \nodata                      & $\rho_{g0,1}=2.1\pm 8.0^{\ast}$ & $\rho_{g,c}=0.82\pm 0.10^{\ast}$                          &     \nodata                                & $\alpha_1=0.21\pm0.21$              \\ 
  \nodata                      & $\rho_{g0,2}=1.1\pm 0.1^{\ast}$                                & \nodata                           &     \nodata                                & $\alpha_2=-0.09\pm0.15$             \\
  \nodata                      & \nodata                               & \nodata                           &     \nodata                                & $\gamma=0.45$ (fixed)               \\
  $\chi^2/dof=27.6/4$          & $\chi^2/dof=0.67/2$                   & $\chi^2/dof=1.08/3$               &     $\chi^2/dof=21.3/5$                    & $\chi^2/dof=15.0/3$                 
\enddata
\tablecomments{Models of the gas density and temperature as a function
of three-dimensional radius. The functions and parameters are defined
in \S \ref{density} and \S \ref{temp}. The gas density normalizations
($^{\ast}$) are expressed in units of $10^{-26}$ g cm$^{-3}$.}
\end{deluxetable*}

Our analysis of the gravitating mass and dark matter requires that we
evaluate derivatives of the gas density and temperature with respect
to the three-dimensional radius (\S \ref{gravitating_mass} and \S
\ref{dm}). To reduce the noise in the derivative calculations we fit 
simple analytic functions to the entire radial range of the gas
density and temperature data. For the gas density ($\rho_g(r)$) we
integrate the quantity, $\rho_g^2\Lambda(T,\fe)$, along the
line-of-sight, where $\Lambda(T,\fe)$ is just the \apec\ model with
$norm=1$. The temperature profile model is fixed to the best-fitting
result obtained in the following section. (The iron abundance profile
is also fitted with a simple model and extrapolated as a power-law
outside the last data point. It also remains fixed during the fit.) 
The limits of integration are projected radius $R$ and a maximum
value, the latter of which we set near 2~Mpc corresponding to the
inferred virial radius of the cluster below. (The results are not
sensitive to this choice.) Consequently, the integral is obtained as a
function of projected radius $R$ which we then evaluate over the
radial width of each circular annulus defined in Table
\ref{annuli}. The result for each annulus is divided by
$\Lambda(T,\fe)$, where now $T$ and $\fe$ are the emission-weighted
results quoted in Table \ref{annuli} for the annulus in question.

As is standard for such studies, we begin by fitting the gas density
profile with the
well-known $\beta$ model \citep{beta},
\begin{equation}
  \rho_g=\rho_{g0} \Big[1+\Big(\frac{r}{r_c}\Big)^2\Big]^{-\frac{3}{2}\beta} ,
\end{equation}
where $\rho_{g0}$ is the central gas density, $r_c$ the core radius,
and the asymptotic slope is $-3\beta$. The result of fitting this
model to the data is shown in Figure~\ref{density_profile} and the
parameters are listed in Table~\ref{proj_prof}. Although the single
$\beta$-model is not formally an acceptable fit, it provides a good
representation of the radial profile. The fractional residuals are
$<5\%$ in all annuli except the center where a $\sim 10\%$ deviation
is observed.

Formally acceptable fits can be obtained by employing conventional
modifications of the single $\beta$ model. First, we examine the
``cusped $\beta$~model'' \citep{pratt,lewi03a},
\begin{equation}
   \rho_g=\rho_{g,c}2^{3\beta/2-\alpha_c/2}\left(\frac{r}{r_c}\right)^{-\alpha_c}
   \left[1+\left(\frac{r}{r_c}\right)^2\right]^{-\frac{3}{2}\beta+\frac{\alpha_c}{2}}
   ,
\label{cusp_beta}
\end{equation}
where the exponent $\alpha_c$ is the slope of the power-law cusp at
small radii and $\rho_{g,c}=\rho_g(r_c)$. Although this model
introduces only one additional free parameter over the single $\beta$
model, as is seen in Figure~\ref{density_profile} and
Table~\ref{proj_prof} the fit is excellent: residuals at all radii are
reduced to $\la 1\%$.

Second, we also investigated adding a second $\beta$ model; i.e., a
``double-$\beta$'' model \citep[e.g.,][]{xu,mohr}. Even if we require
the $\beta$ values of both components to be the same, the fit is
excellent -- just as good as the cusped $\beta$~model (see
Figure~\ref{density_profile} and Table~\ref{proj_prof}). Since this
model does not improve the fit over the that achieved by cusped
$\beta$~model, but it introduces another free parameter, we will use
the cusped $\beta$~model as the default profile in the mass analysis
below.

The values we obtained for $\beta$ are larger than we obtained
previously for A2589 with \chandra\ data \citep{buot04a}. Using a
single $\beta$ model, which was all that was required by the
low-quality ACIS-S observation, we obtained $\beta=0.39\pm 0.04$;
i.e., $\approx 4\sigma$ smaller than the result obtained for the single
$\beta$ model with the EPIC data. This difference is not surprising
since the \chandra\ data were only fitted out to $\approx 150$~kpc
compared to $\approx 500$~kpc in this paper. The $\beta$ values
obtained from \xmm\ agree well with the value of 0.57 quoted by
\citet{davi96a} obtained from {\sl ROSAT} data.

\section{Gas Temperature profile}
\label{temp}

\begin{figure}[t]
  \begin{center} \includegraphics[width=0.5\textwidth,
  angle=0]{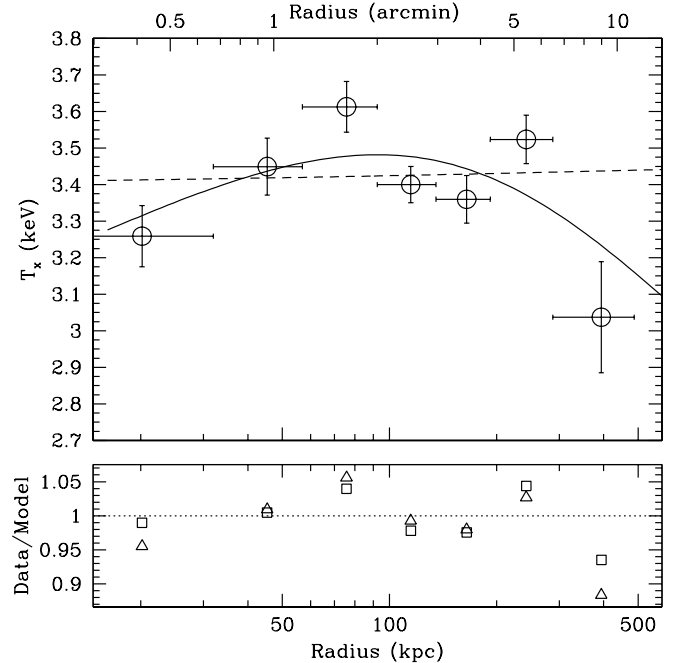} \caption{Projected temperature
  profile. Open circles represent our \xmm\ observation. The models
  are emission-weighted projections averaged over each annulus (see \S
\ref{temp}). However, for display purposes we do not bin the models
in the plot.  The solid line is the emission weighted $T(r)^*$
model. The dashed line is the emission weighted power-law. In the
ratio plot the triangles represent the power-law and squares are the
$T(r)^*$ model.}  \label{temp_profile} \end{center}
\end{figure}

We display the temperature profile in Figure
\ref{temp_profile} corresponding to the values in Table
\ref{annuli}. The approximately isothermal profile is similar 
to that measured by \chandra\ in \citet{buot04a} and does not show the
evidence for a cool core indicated by \rosat\ \citep{davi96a}.  The
lack of a cool core is unusual for a cluster having such a regular
X-ray morphology, though the centrally peaked iron abundance profile
is similar to those found in cool core clusters
\citep[e.g.,][]{degr04a,bohr04a}.

Starting with a model for $T(r)$ we projected the quantities
$T\rho_g^2\Lambda(T,\fe)$ and $\rho_g^2\Lambda(T,\fe)$ along the line
of sight. (Here we use the best-fitting model for $\rho_g(r)$ obtained
above, and the \fe\ profile is fixed as before.) These quantities were
evaluated within annuli in the same manner as done for the gas
density. The projected emission-weighted temperature is obtained by
dividing the first term by the second term.

Since the temperature data are nearly isothermal, we initially fitted
a single power-law,
\begin{equation}
\label{pow_eq}
T(r) = T_{100} \Big(\frac{r}{100\,\rm{kpc}}\Big)^{\alpha_p}  ,
\end{equation}
where $T_{100}$ is the temperature at 100~kpc. We show the
best-fitting model in Figure \ref{temp_profile} and list the
parameters in Table \ref{proj_prof}. The power-law exponent is
$\alpha_p=0.00\pm 0.02$, indicating an isothermal profile.  This
result is also consistent with that obtained for the
\chandra\ data by \citet{buot04a}; the normalization of the
temperature profile obtained from the \xmm\ data is also within
$1.2\sigma$ of that obtained from \chandra.

Unlike the low-quality \chandra\ temperature profile, the \xmm\ 
temperature data are not formally consistent with the isothermal /
power-law model. A better fit can be achieved by adding another degree
of freedom to account for the curvature of the profile in the log-log
plot. We experimented with several models and adopted a model
consisting of two power-laws joined by an exponential cut-off term,
\begin{equation}
\label{pow2expcut2}
T(r)^{*} =  T_{1}\, e^{-(\frac{r}{r_p})^{\gamma}}  + T_{2}\, (1 - e^{-(\frac{r}{r_p})^{\gamma}})  ,
\end{equation}
where $T_{i=1,2}=A_{i}\Big(\frac{r}{r_p}\Big)^{\alpha_i}$ and $A_i$,
$r_p$, $\alpha_i$ and $\gamma$ are respectively the normalization,
scale radius and slope for the power-laws and the exponent of the
power-law functions in the exponentials. This model provides a better
fit (Figure \ref{temp_profile} and Table \ref{proj_prof}), though only
the fractional residuals of the innermost and outermost data points
are affected substantially.

We adopt this profile as our default in our subsequent analysis. In
\S~\ref{systematics_sect} we assess how different choices of
temperature profiles affect our mass measurements. There we also
assess the importance of the last data point for determining the best
model. 

\section{Mass analysis}
\subsection{Gravitating mass}\label{gravitating_mass}

\begin{figure}[t]
  \begin{center} \includegraphics[width=0.5\textwidth,
  angle=0]{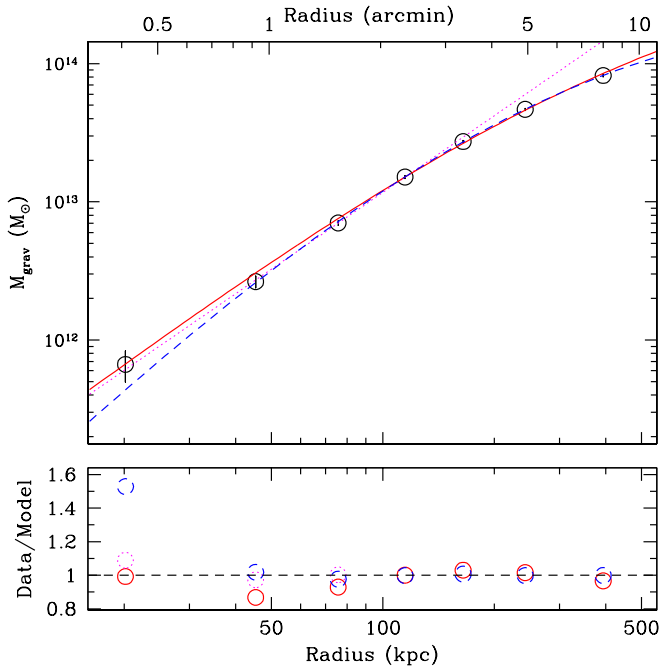} \caption{Gravitating mass
  profile obtained using the cusp $\beta$ model for the gas density
  and the $T(r)^{*}$ model for the temperature. The gravitating mass
  is evaluated at radii corresponding to the annuli used for spectral
  analysis (\S \ref{spec_analysis}). We display the best-fitting NFW
  (solid line), N04 (dashed line) and power-law (only for the three
  inner data points; dotted line) models.}  \label{mass_profile}
  \end{center}
\end{figure}

\begin{deluxetable*}{@{\extracolsep{\fill}}lccccccccc}
\tablecaption{Mass profile fit parameters\label{mass_pars}}
\tabletypesize{\scriptsize}
\tablehead{
Profile     &$\chi^2_c\pm\sigma_{\chi^2}/dof$ &   $r_s \rm{(kpc)}$      &   c        & $\alpha$      &   $r_{vir} \rm{(Mpc)}$         &    $M_{vir} (10^{14}\rm{M_\odot})$ & $c_{200}$    & $M_{200} (10^{14}\rm{M_\odot})$
}
\startdata
NFW         &$ 12.7\pm7.3/5  $                &      $286\pm27$      & $6.1\pm0.3$  & \nodata        &   $1.74\pm0.05$     	   & $3.25\pm0.29$                      & $4.6\pm0.3$  & $2.65\pm0.20$\\      
N04         &$\;\;2.4\pm1.9/4$   	      &      $206\pm21$      & $7.0\pm0.3$  & $0.40\pm0.05$ &   $1.44\pm0.08$              & $1.84\pm0.33$                      & $5.5\pm0.3$  & $1.70\pm0.24$\\      
Power-law   &$\;\;0.2\pm0.2/1$                &        \nodata        &  \nodata      & $1.84\pm0.18$ &     \nodata                   & $0.12\pm0.01$                      & \nodata     &   \nodata    
\enddata
\tablecomments{For the N04 model $r_s$ corresponds to $r_{-2}$. For
the power-law model $\alpha$ corresponds to $\alpha_p$ and the
$M_{vir}$ value is the normalization of the model at
$100\,\rm{kpc}$. The Power-law model is fitted only to the inner
three points.}
\end{deluxetable*}

When hydrostatic equilibrium is a suitable approximation for the hot
ICM the gravitating mass ($M_{\rm grav}$) enclosed within a certain
radius $r$ can be inferred from the density and temperature of the
emitting plasma \citep{mathews,fabricant}:
\begin{equation}
\label{mtot_1}
  M_{\rm grav}(<\!r)  = \frac{k_B}{G\mu m_p}\,rT\left(-\frac{d\ln \rho}{d\ln r}-\frac{d\ln T}{d\ln r}\right)  ,
\end{equation}
where $k_B$ is the Boltzmann's constant, $G$ is the constant of
gravitation, $\mu$ is the mean atomic weight of the gas (taken to be
0.62), $m_p$ is the atomic mass unit.  We adopt the cusp $\beta$ model
for the gas density and the $T(r)^{*}$ model for the temperature for
our fiducial analysis.  Using these parameterizations we evaluate the
gravitating mass at weighted radii, $r\equiv [(r_{\rm out}^{3/2} +
r_{\rm in}^{3/2})/2]^{2/3}$ \citep[see][]{lewi03a} within the annuli used
for spectral analysis (\S \ref{spec_analysis}). The resulting mass
data points are displayed in Figure~\ref{mass_profile}. The error bars
on the data points are the standard deviations derived from
the gas density and temperature profiles obtained from each of the 20
Monte Carlo error simulations (\S \ref{spec_analysis}). Because the
mass data points are correlated, it is inappropriate to use the
$\chi^2$ null-hypothesis probability assuming uncorrelated errors to
assess goodness-of-fit. Instead for each mass model we provide the
standard deviation of $\chi^2$ obtained from fitting a particular mass
model (e.g., NFW) to the mass profile obtained for each of the 20
Monte Carlo simulations. This allows one to assess the relative
goodness-of-fit between different mass models.

First we examined whether the NFW profile \citep{nfw} could provide a
good description of the mass data. The NFW density profile is given by: 
\begin{equation}
  \rho(r)=\frac{\rho_c(z)\delta_c}{(r/r_s)(1+r/r_s)^2} ,
\label{nfw}
\end{equation}
where $\rho_c(z)$ is the critical density of the universe at redshift $z$, 
$\delta_c$ is a characteristic dimensionless density defined as: 
\begin{equation}
  \delta_c = \frac{1}{3}\frac{\rho_{vir}}{\rho_c(z)}\frac{c^3}{\ln(1+c)-c/(1+c)} ,
\end{equation}
where $r_s$ is the scale radius (the point where the logarithmic slope
reaches a value of -2), $\rho_{vir}$ is the mean density of a sphere
enclosed in $r_{vir}=cr_s$ (where $c$ is the concentration parameter)
that is $104.7\rho_c(z)$ \citep[see][]{bryan}. By integrating
equation~\ref{nfw} we obtain the expression of the total mass enclosed
within a radius $r$:
\begin{equation}
  M_{NFW}(<r)=A_{NFW} \left[\ln(1+r/r_s) - \frac{r/r_s}{1+r/r_s} \right] ,
\end{equation}
where $A_{NFW}=4\pi\rho_c(z)\delta_c r_s^3$.  A distinctive feature of
the NFW profile is that the logarithmic slope of the density profile asymptotes
to -1 at small radius corresponding to a logarithmic slope in the mass
of 2. 

We plot the best-fitting NFW model in Fig.~\ref{mass_profile} (solid
line) and list the derived parameters in Table~\ref{mass_pars}. This
fit is very good in the sense that the fractional residuals are $<5\%$
for the central radial bin and for $r>100$~kpc. The largest deviation
is observed for $r\sim 50$~kpc where the fractional residual is $\sim
13\%$. It is interesting that this radius corresponds to the largest
center shift in the X-ray isophotes (\S \ref{image_analysis}).  The
values of $c$ and $\mvir$ obtained for the NFW fit are very consistent
with those obtained from recent observations of relaxed clusters with
\chandra\ and \xmm\ \citep[e.g.,][]{lewi03a,poin05a,vikhlinin06}. (These
results are also consistent with those we obtained previously from the
low-quality \chandra\ AO-3 data.)  These $c$ and $\mvir$ values agree
extremely well with the mean relation predicted by the $\Lambda$CDM
simulations of \citet{bull01a}, where we have extended their toy model
up to virial masses appropriate for A2589.

To obtain an estimate of the slope of the inner density profile
without reference to the NFW model we fitted the three inner mass data
points with a power-law (see Eq.~\ref{pow_eq}; thin dotted line in the
plot). The value we obtain is $1.84\pm0.18$ for the mass which
corresponds to $-0.84\pm0.18$ for the density. This value is
consistent with the inner logarithmic slope given by an NFW profile
within the $1\sigma$ error. The inner slope of -0.8 is also consistent
with a direct solution of the radial Jeans equation for DM assuming an
isotropic velocity dispersion tensor and a CDM phase-space density
\citep{hans05a}.

Finally, we also examined the Sersic-like profile \citep[][hereafter
N04]{nava04a} that was recently suggested as a better
parametrization for CDM halos, particularly within the very inner
regions (less then few percent of the virial radius) where it is
typically shallower than NFW \citep[see also ][]{grah05a}. The N04
density profile is defined as follows,
\begin{equation}
  \rho(r) = \rho_{-2}\, \rm{exp\left({\frac{2}{\alpha}}\right) exp\left[{-\frac{2}{\alpha}x^{\alpha}}\right]} ,
\end{equation}
where $x=r/r_{\rm{-2}}$, and the quantities flagged with -2 refer to
the point where the profile reaches a logarithmic slope of -2
($r_{\rm{-2}}$ is the analog of $r_s$ in the NFW profile). The
$\alpha$ exponent determines the bend of the profile about
$r_{\rm{-2}}$. The enclosed mass obtained by integrating this
expression is:
\begin{equation}
M_{N04}(<r)=A_{N04} \,\frac{1}{\alpha}\,\rm{exp\bigg(\frac{2}{\alpha}\bigg)} \bigg(\frac{\alpha}{2}\bigg)^{3/\alpha} \gamma\bigg(\frac{3}{\alpha},\frac{2}{\alpha}x^{\alpha}\bigg) ,
\end{equation}
where $A_{N04}=4\pi\rho_{-2}r_{-2}^3$ and
$\gamma(\eta,\lambda)=\int_0^{\lambda} t^{\eta-1} e^{-t} dt$ is the
lower incomplete gamma function.

We display the best-fitting N04 profile in Figure~\ref{mass_profile}
represented by a dashed line. The fractional residuals are smaller
than the NFW profile at all radii except the central data point where
the residual is about 50\%. The smaller $\chi^2$ value obtained for
the N04 arises primarily from the better fit of the data point $r\sim
50$~kpc. The inferred value of $\alpha=0.40\pm0.05$ is quite large and
incompatible with the mean value of $0.172\pm0.032$ for CDM halos
\citep{nava04a}. Consequently, the result we obtain for the
concentration and virial mass using the N04 model must be interpreted
with caution; i.e., the large $\alpha$ implies a density profile that
is shallower in the center and steeper at large radii than CDM.

Using our measurement of the gravitating mass obtained from the NFW
model and the gas mass obtained by integrating the cusp $\beta$ model
we compute the gas fraction as a function of radius. For the central
data point we obtain $f_{gas}=0.014\pm0.004$ while it rises to
$0.083\pm0.002$ for our outermost mass data point. This value is
consistent with values obtained by
\citet{sanderson} and \citet{vikh05a} for clusters of similar
temperature. If we extrapolate our fits to the virial radius we obtain
a total gas fraction $f_{gas}=0.168\pm0.009$, consistent with the
value of the baryon fraction $\Omega_b/\Omega_m\sim0.17$ measured by
WMAP \citep{spergel}.

\subsection{The Ratio of Gravitating Mass to cD Stellar Light}

The radial profile of the ratio of gravitating mass to stellar light
in the cluster provides a useful diagnostic for the radius where dark
matter dominates the cluster mass budget. We shall only consider the
optical light of the cD galaxy NGC~7647 since only sparse information
in the literature exists for a small number of other member
galaxies. Fortunately, the $V$-band surface brightness profile of the
cD galaxy has been analyzed by \citet{malumuth} out to a radius of
$\sim130\,\rm{kpc}$. They find that a King model is a good fit to the
optical data with a core radius, $r_c=0.98\,\rm{h_{70}^{-1} kpc}$, and
a total $V$-band luminosity $L_V = 2.3\times
10^{11}\,\rm{L_{V,\odot}}$ within the region analyzed. They find that a de
Vaucouleurs model with $r_e=32.6$~kpc fits the optical data nearly as
well. Therefore, we represent $L_V(r)$ by a Hernquist model
\citep{hern90} with scale radius $a=r_e/1.8153=18.0$~kpc and total
luminosity quoted above, since the projected Hernquist profile is a
good approximation to a de Vaucouleurs profile.

\begin{figure}[t]
  \begin{center} \includegraphics[width=0.5\textwidth,
  angle=0]{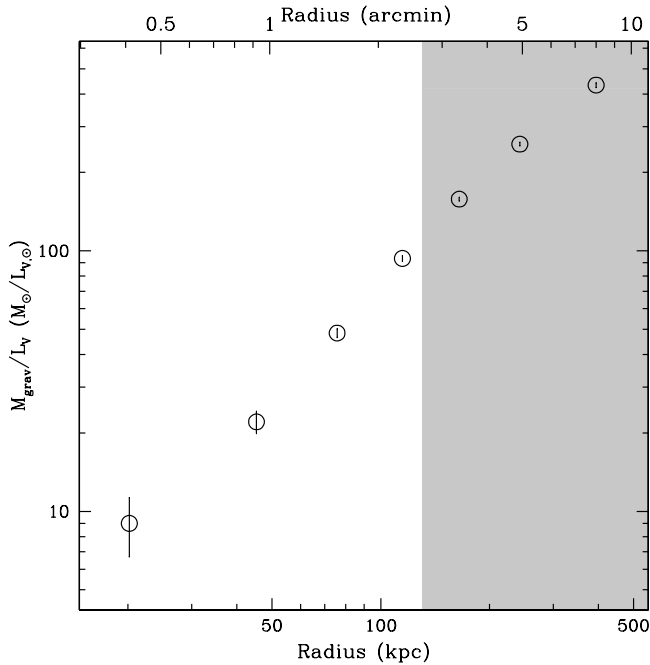} \caption{Ratio of the total gravitating
  mass to the optical light of the cD galaxy. The data in the gray
  area are estimated from an extrapolation of the Hernquist profile
  approximation of the De Vaucouleurs profile obtained by
  \citet{malumuth} to map the V-band light profile out to $\sim
  130\,\rm{kpc}$.}  \label{m2l} \end{center}
\end{figure}

In Fig.~\ref{m2l} we plot $M_{grav}(<\!r)/L_V(<\!r)$: the ratio of
gravitating mass, computed using the NFW model, to the integrated
V-band luminosity of the cD, using the Hernquist model, as a function
of radius.  The $M_{grav}/L_V$ profile increases from
$9.0\pm2.3\,\rm{M_\odot/L_{V,\odot}}$ for the inner data point to
$93.5\pm2.3\,\rm{M_\odot/L_{V,\odot}}$ at the radius representing the
outer extent of the optical data. If we extrapolate the Hernquist
model of the cD out to the extent of the X-ray measurements,
$M_{grav}/L_V$ increases to $433\pm9\,\rm{M_\odot/L_{V,\odot}}$ (gray
area in Fig.~\ref{m2l}).  This latter value is an upper limit to the
cluster mass to light ratio because we have not considered additional
contributions from other member galaxies.

There is no evidence that the $M_{\rm grav}/L_V$ profile flattens at
small radius such as we have observed for A2029
\citep{lewi03a} and for a sample of early-type galaxies
\citep{humphrey06b}.  Since the flattening in the profiles in other
systems occurs for $r\la 10$~kpc, yet our central bin has outer radius
$r=32$~kpc, we do not expect to detect this effect. High quality
\chandra\ data would allow $M_{\rm grav}/L_V$ to be probed on
the small scales necessary for an interesting comparison to other
systems.

We mention that \citet{malumuth} used their fit of the King model and
a measurement of the central velocity dispersion to estimate $M_{\rm
grav}/L_V$ at the very center; i.e., within the King core radius
($<1$~kpc). They obtain $M_{grav}/L_V\approx 15$ in solar units,
converting their value to our redshift and cosmology. This value is a
factor of 1.7 larger than we have measured at $r\sim 20$~kpc, although
the discrepancy is not highly significant ($2.6\sigma$). For a single
burst stellar population with age ranging from 10-13~Gyr and
metallicity ranging from 0.5-2~solar the $V$-band stellar
mass-to-light ratio is only expected to take values from 5-10 in solar
units, with values at the upper end assuming a Salpeter IMF. (We have
used the results of \citealt{mara98a} from updated model grids
provided by the
author.\footnote{http://www-astro.physics.ox.ac.uk/$\sim$maraston/
Claudia's\_Stellar\_Population\_Models.html}) The gravitating mass we
measure for $r\sim 20$~kpc is similar to that we have measured at that
radius in the cluster A2029 and a sample of elliptical galaxies
\citep{lewi03a,humphrey06b}. If the difference in masses estimated from
the X-ray and optical methods is real, it likely implies the stellar
mass-to-light ratio in the cD varies with radius.

\subsection{Dark matter}\label{dm}

We desire to extract the radial profile of the DM from the
gravitating matter by removing the contributions of the known luminous
mass components. Since the X-ray data provide a direct measurement of
the hot gas density, it is straightforward to compute the radial
profile of $M_{\rm grav} - M_{\rm gas}$. Consequently, as a first step
we fitted single-component models to this ``DM'' profile
analogously to our procedure in the previous section. 

The fitted parameters for the NFW, N04, and power-law models are
listed in Table~\ref{dm_pars}. Since the hot gas contributes $<10\%$
to the total mass over the radial range fitted, it is not surprising
that the parameters we infer are very consistent with those obtained
when fitting the total gravitating matter (Table~\ref{mass_pars}).
However, we emphasize that \rvir\ and \mvir\ (and hence $c$) obtained
in this way refer only to the DM component. To obtain the true values
that include the mass of the hot gas we add $M_{\rm gas}$ to $M_{\rm
DM}$, compute a new \rvir, and iterate. For the NFW model this yields
$\rvir=1.77\pm 0.04$~Mpc, $\mvir=3.46\pm 0.26\,
\times 10^{14}\msun$, $c=6.8\pm 0.4$, and a total gas fraction,
$f_g=0.16\pm 0.01$. It should be remembered that these re-scaled
values depend on the extrapolation of our model for $\rho_g$ out to
\rvir.

Next we attempted to better isolate the DM profile by removing the
contribution of the mass from the stars in the cD. To model the cD
stars we employed the Hernquist parameterization of the $V$-band
luminosity described in the previous section, which we now refer to as
the H90 model. We fitted two-component models, DM + stars, to the
radial profile of $M_{\rm grav} - M_{\rm gas}$, where the total mass
in the stars is specified by the stellar mass-to-light ratio,
${M_*/L_{V}}$.

The results of the fits are listed in Table~\ref{dm_pars}. If
${M_*/L_{V}}$ is allowed to be a free parameter in the NFW+H90 fit,
then an unphysically small value ${M_*/L_{V}}\sim 0.15$ (solar
units) is obtained.  The NFW model itself is all that is required to
describe the mass profile; adding in a separate stellar component only
degrades the fit.

\begin{figure*}[!t]
  \begin{center}
    \includegraphics[width=0.49\textwidth, angle=0]{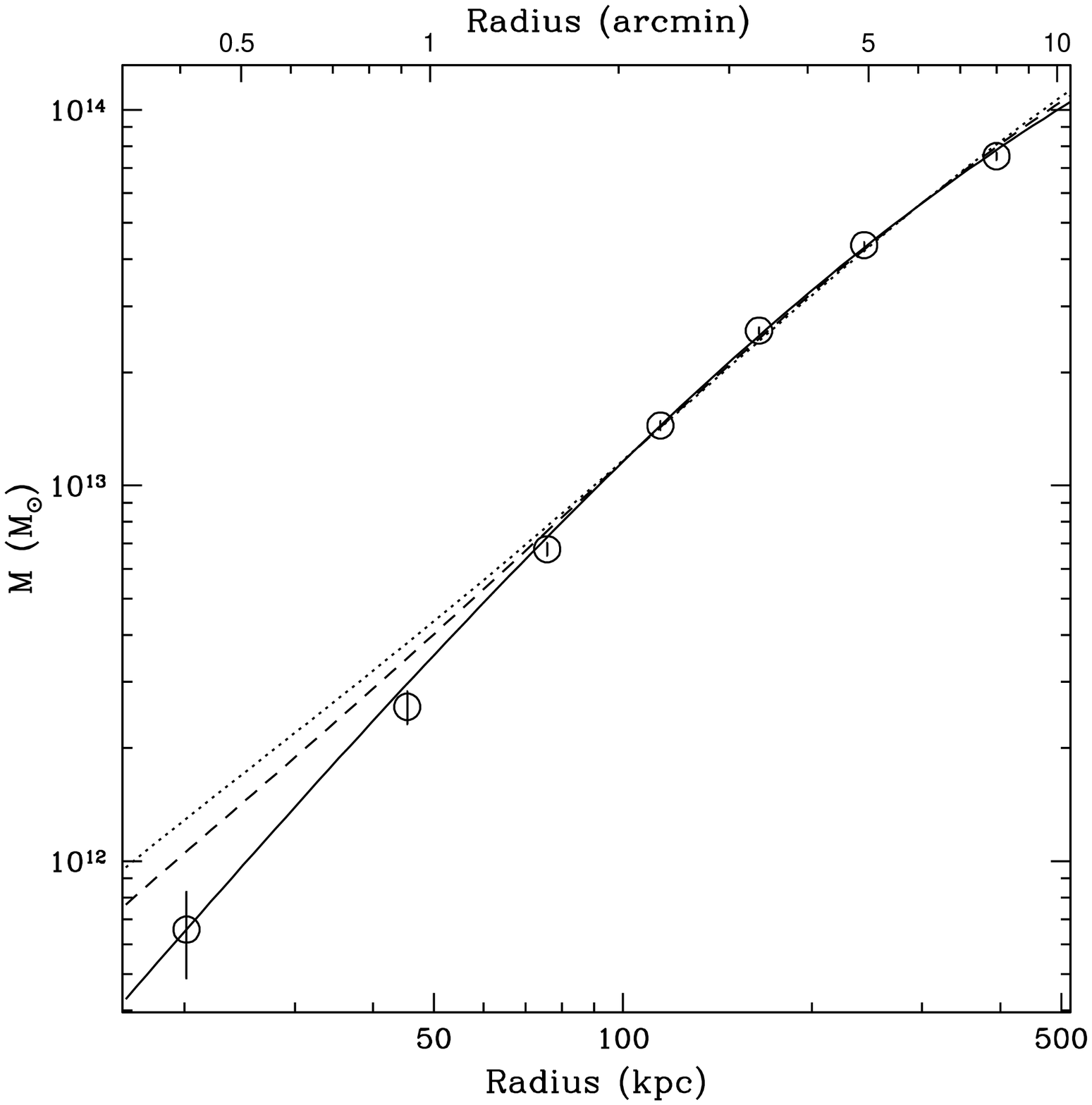}
    \includegraphics[width=0.49\textwidth, angle=0]{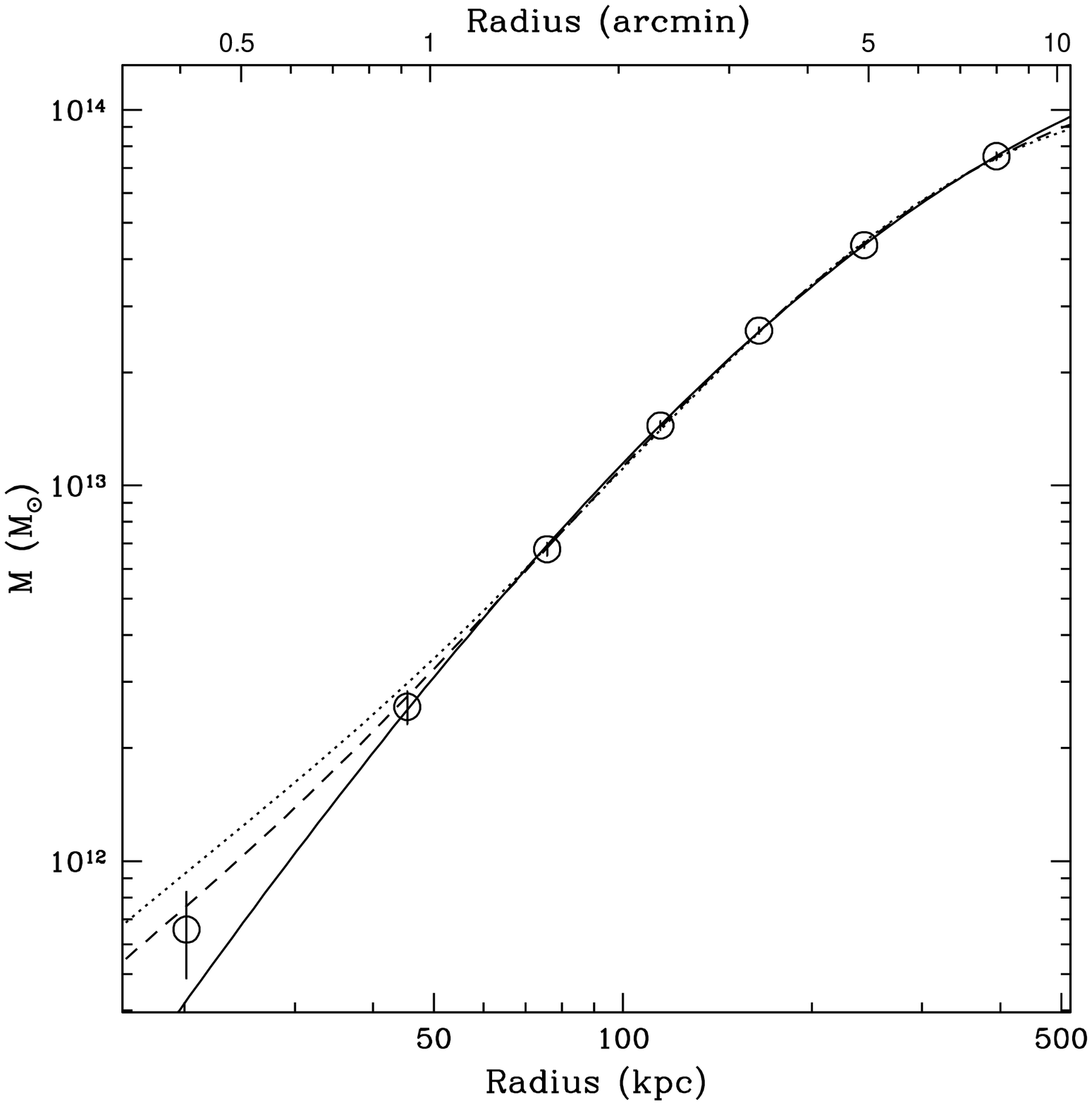}
    \caption{Models fitted to $M_{\rm grav}-M_{\rm gas}$. ({\em left
panel}) The solid line represents the pure NFW profile, the dashed
line adds the stellar component (NFW+H90) with $M_{*}/L_V=9.0$ in
solar units, and the dotted line applies adiabatic contraction to the
NFW profile (NFW*AC+H90) with $M_{*}/L_V=9.0$. ({\em right
panel}) Same as left panel except the NFW profile is replaced with the
N04 profile.}
  \label{dmstars_profile}
  \end{center}
\end{figure*}

We desired to explore the importance of adiabatic contraction of the
DM halo arising from the early condensation of baryons in the central
galaxy \citep[e.g.,][]{blum86a,gned04a}. We refer to the modified NFW
fit as ``NFW*AC+H90'', indicating that adiabatic contraction only
applies to the DM component\footnote{The adiabatic contraction code we
used was made publicly available by Oleg Gnedin at:
http://www.astronomy.ohio-state.edu/$\sim$ognedin/contra/}. Because
the fitted ${M_*/L_{V}}$ is so small, the AC model does not have a
substantial effect when ${M_*/L_{V}}$ is allowed to be a free
parameter. Consequently, we also tried fixing
${M_*/L_{V}}=9.0$~(solar units) corresponding to the gravitating
mass-to-light ratio obtained for our inner data point, as well as
being a reasonable estimate from stellar population synthesis models,
as discussed in the previous section. With this ${M_*/L_{V}}$ the
NFW+H90 and NFW*AC+H90 fits are quite poor within $r\sim 50$~kpc, and
the effect of adiabatic contraction on the derived parameters for the
DM is roughly equal to that arising from adding the stellar component;
e.g., $c$ changes from 6.3 (NFW) to 5.2 (NFW+H90) and from 5.2
(NFW+H90) to 4.3 (NFW*AC+H90). The adiabatic contraction of the dark
matter produces a more centrally concentrated halo that deviates even
more strongly with the data within $r\sim 50$~kpc (see Figure
~\ref{dmstars_profile}).

Since the N04 model fitted to the total gravitating matter in \S
\ref{gravitating_mass} underestimates the mass at the center, it is
expected that, unlike the NFW profile, it will allow for a substantial
contribution from a central stellar mass component. Indeed, the
N04+H90 fit yields a stellar mass-to-light ratio, ${M_*/L_{V}}=
4.8\pm 1.9$, which is consistent with expectations of single-burst
stellar populations synthesis models. Applying adiabatic contraction
(i.e., N04*AC+H90) yields a smaller, though still reasonable, value,
${M_*/L_{V}}= 3.1\pm 0.7$. However, the values of $\alpha\sim 0.5$
obtained for these models, even larger than obtained for the fit to
the gravitating matter, are even more inconsistent with the values of
$\sim 0.20$ required to represent CDM halos \citep{nava04a}. Also,
exactly as for the total gravitating matter, it is the data point near
$r\sim 50$~kpc that is largely responsible for the better fit of the
high-$\alpha$ N04 model with respect to NFW (see Figure
~\ref{dmstars_profile}).

\begin{deluxetable*}{@{\extracolsep{\fill}}lccccccc}
\tablecaption{Model Fits to $M_{\rm grav} - M_{\rm gas}$ \label{dm_pars}}
\tabletypesize{\scriptsize}
\tablehead{
\multicolumn{8}{c}{Single-Component Models: Dark Matter Only}\\
Model                  &    $M_*/L_V$  &   $\chi^2_c\pm\sigma_{\chi^2}/dof$ &   $r_s \rm{(kpc)}$   &   c           & $\alpha$         &   $r_{vir} \rm{(Mpc)}$    &    $M_{vir} (10^{14}\rm{M_\odot})$       
}
\startdata 								    
NFW                               & \nodata       &$ 12.4\pm7.1/5   $    &   $264\pm26$         & $6.3\pm0.4$   & \nodata           &   $1.65\pm0.05$                & $2.78\pm0.25$         \\ 
N04                               & \nodata       &$\;\;2.4\pm1.9/4 $    &   $193\pm19$         & $7.1\pm0.3$   & $0.41\pm0.07$    &   $1.37\pm0.08$                & $1.58\pm0.28 $         \\ 
Power-Law                         & \nodata       &$\;\;0.2\pm0.2/1 $    &   \nodata             & \nodata        & $1.82\pm0.18$    &   \nodata                       & $0.11\pm0.01$        \\ 
\hline
\\[0.1cm]
\hline
\hline
\relax\\[-1.7ex]%
\multicolumn{8}{c}{Two-Component Models: Dark Matter + cD Stars}\\
Model                  &    $M_*/L_V$  &   $\chi^2_c\pm\sigma_{\chi^2}/dof$  &  $r_s \rm{(kpc)}$   &   c           & $\alpha$         &   $r_{vir} \rm{(Mpc)}$    &    $M_{vir} (10^{14}\rm{M_\odot})$       \\
\relax\\[-2.2ex]%
\hline 
\relax\\[-1.5ex]%
NFW+H90                           & $(9)          $  & $42.0\pm18.0/5   $              & $  338\pm21     $  & $5.2\pm0.3$     &  \nodata           &   $1.74\pm0.05 $          & $3.26\pm0.29$   \\      
NFW+H90                           & $0.15\pm0.02  $  & $12.7\pm7.1/5    $              & $  263\pm16     $  & $6.2\pm0.3$     &  \nodata           &   $1.65\pm0.05 $          & $2.78\pm0.25$   \\      
NFW*AC+H90                        & $(9)          $  & $76.9\pm25.2/5   $              & $  428\pm32     $  & $4.3\pm0.3$     &  \nodata           &   $1.84\pm0.07 $          & $3.84\pm0.45$   \\      
NFW*AC+H90                        & $0.055\pm0.001$  & $12.6\pm7.2/4    $              & $  264\pm18     $  & $6.2\pm0.4$     &  \nodata           &   $1.65\pm0.05 $          & $2.78\pm0.25$   \\
N04+H90                           & $(9)          $  & $\;\;2.0\pm4.2/4 $              & $\!\!190\pm\;6  $  & $6.6\pm0.1$     & $0.61\pm0.05$      &   $1.25\pm0.03 $          & $1.20\pm0.09$   \\      
N04+H90                           & $4.8\pm1.9    $  & $\;\;0.7\pm0.7/3 $              & $ \!\!190\pm\;9 $  & $6.8\pm0.2$     & $0.51\pm0.05$      &   $1.30\pm0.04 $          & $1.34\pm0.12$  \\
N04*AC+H90                        & $(9)          $  & $\;8.2\pm7.1/4   $              & $\!\!195\pm\;6  $  & $6.1\pm0.1$     & $0.81\pm0.08$      &   $1.20\pm0.03 $          & $1.04\pm0.07$   \\     
N04*AC+H90                        & $3.1\pm 0.7   $  & $\;0.8\pm1.1/3   $              & $ \!\! 191\pm\;7$  & $6.8\pm0.1$     & $0.53\pm0.06$      &   $1.29\pm0.04 $          & $1.32\pm0.13$
\enddata
\tablecomments{For the power-law model $M_{vir}$ is the value of the
normalization at $100\,\rm{kpc}$. Fixed parameters are enclosed in
parenthesis. ``*AC'' indicates the dark matter component is compressed
adiabatically following the prescription of \citet{gned04a}. Note that
the virial radii and mass refer only to the dark matter component.}
\end{deluxetable*}

\renewcommand{\arraystretch}{1.25}
\begin{deluxetable*}{@{\extracolsep{\fill}}lcccccccc}
       \tablecaption{Estimation of the systematic errors\label{systematics}}
\tabletypesize{\scriptsize}
\tablehead{
         \multicolumn{9}{c}{NFW*AC+H90}\\
         Parameter      & $\rm{Best fit}$ & $\Delta_{stat}$ & $\Delta_{bkg mod}$ & $\Delta_{bndwidt}$  & $\Delta_{mod} (T/\rho)$ & $\Delta_{exclpts}($1\,-\,6$/$2\,-\,7$)$ & $\Delta_{cen}$ & $\Delta_{deproj}$ 
}
\startdata
         c              & $  6.2    $  &  $ \pm0.4        $  &  $   +0.4\pm0.4     $  &  $    -0.7             $  &  $  +0.6\pm0.6/-1.0                $  &  $    -0.6/+0.6                $  &       \nodata           &  $    -2.4   $      \\
   $M_{vir}           $ & $  2.78   $  &  $ \pm0.25       $  &      \nodata           &  $    +0.71            $  &  $    -0.30/+1.22                  $  &  $    +0.92/-0.36              $  &       \nodata           &  $    +4.11  $     \\
   $M_{*}/L_V$          & $  0.055  $  &  $ \pm0.001      $  &  $    -0.054        $  &  $    -0.051           $  &  $    -0.044/-0.041                $  &  $  -0.045/-0.043              $  &  $    -0.044         $  &  $   -0.025  $   \\
         \hline                                                                                                                        
\\[0.1cm]
\hline
\hline
\relax\\[-1.7ex]%
         \multicolumn{9}{c}{N04*AC+H90}\\                                               
         Parameter      & $\rm{Best fit} $  &  $ \Delta_{stat} $  &  $ \Delta_{bkg mod} $  &  $ \Delta_{bndwidt}  $  &  $ \Delta_{mod} (T/\rho) $  &  $ \Delta_{exclpts}($1\,-\,6$/$2\,-\,7$) $  &  $ \Delta_{cen} $  &  $ \Delta_{deproj}$ \\ 
\relax\\[-2.2ex]%
\hline 
\relax\\[-1.5ex]%
         c              & $  6.8    $  &  $ \pm0.1        $  &  $    +0.5          $  &       \nodata                       &  $    \pm0.6/-0.4                  $  &  $   +0.3/+0.5               $  &  $     -0.3           $  &  $    -1.0      $   \\
   $M_{vir}           $ & $  1.32   $  &  $ \pm0.13       $  &  $    -0.21         $  &       \nodata                       &  $   +0.16/-0.10                   $  &  $ -0.50/+0.16\pm0.38        $  &        \nodata           &       \nodata       \\
         $\alpha$       & $  0.53   $  &  $ \pm0.06       $  &  $    +0.07\pm0.05  $  &       \nodata                       &  $   -0.14/+0.34                   $  &  $  +0.26/-0.13              $  &        \nodata           &  $    +0.33     $   \\
   $M_{*}/L_V$          & $  3.1    $  &  $ \pm0.7        $  &  $    -2.4          $  &  $    -0.8\pm1.1                 $  &  $  (-3.1,+4.0)/+2.5               $  &  $+1.5\pm1.2/-3.1            $  &  $     +2.1           $  &  $    +3.3\pm2.2$   
\enddata
\tablecomments{$\Delta_{stat}$ is the $1\sigma$ statistical error
estimate. $M_{vir}$ is expressed in units of
$10^{14}\,\rm{M_{\odot}}$.}
\end{deluxetable*}

\section{Assessment of systematic errors in the mass profile
derivation}\label{systematics_sect}

In this section we provide estimates of the magnitude of systematic
errors on the parameters parameters we have deduced for the dark
matter. We illustrate the effects of systematic errors on two fiducial
two-component models consisting of a stellar component and
adiabatically compressed dark matter: NFW*AC+H90 and N04*AC+H90. The
results are summarized in Table~\ref{systematics}.  The statistical
error of the default model ($\Delta_{stat}$) is also listed in the
table. For each variant model involved in the systematic checks, we
also obtain a corresponding statistical error ($\Delta_{sys}$). If
$\Delta_{sys}>=\Delta_{stat}$ we also quote its magnitude with the
associated best-fitting parameter shift in the table. For example, the
concentration parameter for the NFW*AC+H90 model has a shift of
$0.4\pm 0.4$ in the $\Delta_{bkgmod}$ column; i.e.,
$\Delta_{sys}=\Delta_{stat}$ in this case. We emphasize that none of
the systematic checks that we explored can increase the inferred
$M_{*}/L_V$ for the NFW*AC+H90 model to values 5-10 expected from the
stellar population.

In the following we describe the different checks performed: 
\begin{enumerate}

\item PSF: As stated in \S \ref{spec_analysis} we defined our annuli to
have at least a width of $1\arcmin$; in particular the central radial
bin is a circle of radius $30\arcsec$. With this choice the central
circle encloses about 90\% of the energy of a point source.
Furthermore, since the radial temperature profile is observed to be
nearly isothermal we expect that our measured ``spectroscopic''
temperature should differ negligibly from the emission-weighted
temperature within the bin \citep[e.g.,][]{mazz04a}. The consistency of
our results with those obtained from the lower S/N, but higher
resolution, \chandra\ data (\S \ref{density} and \ref{temp}), and with
\xmm\ and \chandra\ observations of other clusters (see below in
\S \ref{conclusions}), suggests that the larger \xmm\ PSF has not
biased our derived density and temperature parameters by amounts more
than the estimated statistical errors.

 \item Background modeling ($\Delta_{bkg mod}$ in
 Table~\ref{systematics}): the major sources of uncertainty in the
 background modeling are the shape of the broken power-law that takes
 into account the instrumental background at high energies and the
 level of the CXB component. We assess the influence of these
 contributions by a variation of $\pm20\%$ in the slope of the high
 energy part of the broken power-law and in the normalization of the
 CXB component. We also set the parameters of the broken power-law to
 the values obtained fitting the spectrum of the out of field of view
 events. This last check almost always produced the major source of
 uncertainty, especially because it caused the outermost data point to
 vary appreciably. However, this check does not produce results that differ
 appreciably from the best fit value except for the mass to light
 ratio inferred for the N04*AC+H90 profile. \item We varied the
 band-width in which the spectrum in each annulus has been fitted
 ($\Delta_{bndwidt}$ in Table~\ref{systematics}). We choose to
 restrict our analysis to the 0.5-5~keV and 1-5~keV bands. Only the
 NFW profile is affected by this check. 

\item We tried different
 models for the temperature and density profiles (that are shown
 separately in $\Delta_{mod}$ in Table~\ref{systematics}).  The models
 we used are three for the temperature (the first three presented
 below) and one for the density. They are:
\begin{enumerate}
\item the power-law presented in Eq.~\ref{pow_eq};
\item a lognormal profile;
\item a profile that joins smoothly two power-laws: 
  $T(r)=(P_1^{\beta_1}+P_2^{\beta_2})^{-1/\sigma}$, where $P_1$ and
  $P_2$ are the power-laws (Eq.~\ref{pow_eq});
\item a double $\beta$~model with the $\beta$'s tied.
\end{enumerate}
It is worth noting that for both the DM parameterizations the lognormal
profile gives results very similar to the best fit. For the N04*AC+H90
parametrization the third profile does not allow for the presence of
any central stellar component and the power-law temperature model
predict a very high M/L. The different density parametrization has a
noticeable impact on the NFW profile.

\item We tried to exclude from the analysis either the first or the
  last annulus ($\Delta_{exclpts}$ in Table~\ref{systematics}). We
  excluded them either only during the fitting of the dark matter
  profile or also in the temperature and density fitting. We show them
  separately in Table~\ref{systematics} as 1-6 (last point excluded)
  and 2-7 (first point excluded). These points influence critically
  either the form of the dark matter profile (case 1-6) or the
  dominance of the central stellar component (case 2-7).  

\item We
  repeated the analysis using annuli centered on the X-ray centroid of
  the cluster ($\Delta_{cent}$ in Table~\ref{systematics}). The
  parameters did not change much and their differences from the best
  fit values are smaller than the statistical errors.

\item We
  analyzed the effect of the deprojection on the data
  ($\Delta_{deproj}$ in Table~\ref{systematics}) using the
  ``onion-peeling'' technique \citep[e.g.][]{buote2000,buote2003}. The
  quality of the current data does not allow good constraints with
  this technique.

\end{enumerate}

\section{Non-Thermal Pressure Support}\label{nontherm}

Our analysis of the mass profile of A2589 inferred from the \xmm\ data
indicates that the NFW model provides a good description of the total
gravitating mass profile. However, it is widely thought that the
central dark matter profile will be compressed adiabatically because
the stellar baryons would have collapsed at early times
\citep[e.g.,][]{blum86a,gned04a}. As we have shown in \S
\ref{dm} the adiabatic contraction model assuming a reasonable
stellar mass-to-light ratio for the cD galaxy exceeds the mass data in
the central $r\la 50$~kpc.

We investigate whether plausible additional pressure support from
non-thermal processes could reconcile the \xmm\ mass profile with the
adiabatic contraction scenario. Since the most important non-thermal
pressure should arise from turbulent motions and magnetic fields in
the hot ICM, we shall focus our attention on them. We do not consider
possible pressure support from cosmic rays which, in principle, could
be of the same magnitude as that arising from magnetic fields. As
pointed out by \citet{loeb94a}, the fact that S-Z measurements of
clusters are not dominated by a synchrotron signal, the pressure from
cosmic ray electrons should be much less than the thermal pressure in
clusters.  However, the pressure from cosmic ray protons may still be
important.

For simplicity we follow  \citet{loeb94a} and assume,
\begin{equation}
P_{\rm nontherm} = \alpha P_{\rm therm}, \label{eqn.assumption}
\end{equation}
where $\alpha$ is a constant. This expression cannot be strictly valid
since we expect little contribution from non-thermal pressure for
$r\ga 100$~kpc. For our purposes we require only that $d\alpha/dr$ can
be neglected at the radii of interest. If we also assume that the
equation of hydrostatic equilibrium remains a good approximation but
with the thermal gas pressure, $P_{\rm therm}$, replaced by the total
pressure, $P_{\rm nontherm}+P_{\rm therm}$, then we have at any radius
$r$,
\begin{equation}
\alpha = \frac{M(<r)_{\rm NFW*AC+H90}}{M(<r)_{\rm NFW}} - 1, \label{eqn.alpha}
\end{equation}
where $M(<r)_{\rm NFW}$ is the mass determined from fitting the NFW
model to the total gravitating matter (Table \ref{mass_pars}),
and $M(<r)_{\rm NFW*AC+H90}$ is the mass obtained from the adiabatically
contracted NFW model assuming ${M_*/L_{V}}=9.0$ in solar units for
the cD galaxy (Table \ref{dm_pars}).

For our inner mass data point ($r=20.2$~kpc), where the discrepancy is
largest (see Figure \ref{dmstars_profile}), we obtain $\alpha=0.93$
using our best-fitting models. Note that the NFW*AC+H90 value is
$\approx 4\sigma$ larger than the mass data point. For the second mass
data point, $r=45.3$~kpc, we obtain a smaller value, $\alpha=0.27$,
though the NFW*AC+H90 mass value remains $\approx 4\sigma$ larger than
the mass data point. (Note that the NFW mass value is only $1.5\sigma$
larger than the second mass data point.)

\subsection{Turbulence}

If we assume that the only source of non-thermal pressure arises from
random (isotropic) turbulent motions, $P_{\rm turb} =
\frac{1}{3}\rho_g\langle v^2_{\rm turb}\rangle$, then equations
(\ref{eqn.assumption}) and (\ref{eqn.alpha}) imply,
\begin{equation}
v_{\rm turb} = \sqrt{\alpha}\left(\frac{k_BT}{3.26\, \rm
keV}\right)^{0.5}\, \rm 1233\, km\, s^{-1},
\end{equation}
where the gas temperature corresponding to the inner radial bin (Table
\ref{annuli}) has been used. For our inner mass data point
($r=20.2$~kpc) this gives, $v_{\rm turb}\approx 1.3c_s$, where
$c_s=919$~km/s is the adiabatic thermal sound speed of the gas.  The
generally undisturbed appearance of this cluster (\S
\ref{image_analysis}) argues strongly against a turbulent velocity
this large. For the second mass data point, $r=45.3$~kpc, we also have
a very large velocity, $v_{\rm turb}\approx 0.7c_s$, where $c_s =
945$~km/s corresponding to $k_BT=3.45$~keV.  Cosmological simulations
generally find much smaller levels of turbulence in cluster cores,
$v_{\rm turb} = 0.1-0.3 \, c_s$ \citep[e.g.,][]{naga03a,falt05a}. In
fact, for lower mass clusters like A2589, \citet{dola05a} conclude that
only $\approx 5\%$ of the total pressure can be ascribed to turbulent
motions.

A further check on the viability of turbulent pressure support is to
verify whether, $l$, the eddy size corresponding to $v_{\rm turb}$, is
smaller than the length scale under consideration. In an approximate
steady state the energy produced by turbulent motions must be
dissipated on appropriate viscous length scales or else the gas will
be rapidly heated. After setting the rate of turbulent energy
generation equal to the rate of radiation energy loss, we solve for
the eddy size,
\begin{eqnarray}
l & = & \alpha^{1.5}\left(\frac{k_BT}{3.26\, \rm
keV}\right)^{1.5}\left(\frac{\rho_g}{2.6\times 10^{-26}\, \rm g \,
cm^{-3}}\right)^{-1}\times \nonumber \\
   & & \left(\frac{\Lambda(T,Z)}{1.7\times 10^{-23}\, \rm erg \,
cm^{3}\, s^{-1}}\right)^{-1}\, \rm 3.2\, Mpc,
\end{eqnarray}
where $\rho_g$ is the gas density at $r=20.2$~kpc and $\Lambda(T,Z)$ is the
plasma emissivity evaluated for the conditions of the inner annulus (Table
\ref{annuli}). The eddy size is vastly larger than the $r\sim 20$~kpc scale of the central
region under consideration -- and is even larger than the virial
radius of $\sim 1.7$~Mpc. We conclude that pressure arising from
random turbulent motions in the core cannot reconcile our measurement
with the NFW*AC+H90 model.

\subsection{Magnetic Fields}

Now taking magnetic fields to supply all of the necessary non-thermal
pressure ($B^2/8\pi$), we obtain from equations (\ref{eqn.assumption})
and (\ref{eqn.alpha}),
\begin{equation}
B       =  \sqrt{\alpha}\left(\frac{k_BT}{3.26\, \rm
keV} \, \frac{\rho_g}{2.6\times 10^{-26}\, \rm g \,
cm^{-3}}\right)^{0.5}\, \rm 58\, \mu G,
\end{equation}
where, as above, we have taken $T$ and $\rho_g$ appropriate for our
inner mass data point, $r=20.2$~kpc. For $\alpha=0.93$ the required
magnetic field for the inner mass data point is, $B=56\mu$G.  Fields
of similar magnitude have been suggested previously to explain X-ray
mass discrepancies using \rosat\ and \einstein\ data within the
central $<1$~kpc of the disturbed elliptical galaxy NGC~4636
\citep{brig97a}. A field strength of $53\mu$G was also suggested previously
by \citet{loeb94a} to explain the discrepancy between X-ray and
lensing mass estimates in the core of the cluster A2218.

But the typical magnetic fields in galaxy clusters range from
$1-10\mu$G, much lower than required for interesting pressure support
\citep[e.g., for a review see][]{govo04b}. Larger fields have been
measured in some clusters with strong radio sources. A particularly
interesting case is A2029, which, though more massive, appears to be
very relaxed both within the core and on larger scales like A2589
\citep{lewi02a}. Unlike the radio-quiet A2589, the cluster A2029 possesses a
Wide-Angle-Tail radio source from which \citet{eile02a} estimate a
field of $16\mu$G within 8~kpc of the center of A2029.  This field
does not provide interesting pressure support, and the strength likely
declines rapidly with increasing radius \citep{govo01a}.

Although we recognize that there remain outstanding issues in the
determinations of magnetic fields in clusters, current evidence
clearly does not favor field strengths of $\approx 56\mu$G in galaxy
clusters. Therefore, we believe that magnetic pressure support is an
unlikely explanation to reconcile the NFW*AC+H90 model with the
\xmm\ mass data we have presented for A2589. 

\section{Conclusions}\label{conclusions}

Using a new \xmm\ observation we have presented an analysis of the
radial mass profile inferred from the properties of the hot ICM of the
radio-quiet galaxy cluster A2589. We have confirmed the highly regular
X-ray image morphology indicated by previous X-ray observations
possessing lower resolution and/or lower S/N
\citep{davi96a,buot04a}.  The only notable deviation of the X-ray
image from concentric ellipses is a $\approx 16$~kpc shift of the
X-ray center of the $R=45-60$~kpc annulus. The radial temperature
profile is nearly isothermal unlike most X-ray regular clusters which
display cool cores \citep[e.g.,][]{degr02a}. However, the metallicity
does peak at a value near solar at the center and falls off with
increasing radius similar to that in cool-core clusters
\citep[e.g.,][]{degr04a,boehringer}. Overall, given the highly regular
nature of the X-ray image and spectral properties, and the fact it is
bright, nearby, and possess no significant radio emission from the
central galaxy, A2589 is an especially good target for X-ray studies
of its mass distribution.

We find that the NFW profile fits the total gravitating matter
($M_{\rm grav}$) very well over the region studied
($r=0.015-0.25\,\rm{r_{vir}}$, fractional residuals $\lesssim 10\%$)
with $c_{\rm vir}=6.1\pm 0.3$ and $M_{\rm vir} = 3.3\pm 0.3 \times
10^{14}M_{\odot}$ ($r_{vir} = 1.74\pm 0.05$~Mpc) in excellent
agreement with the $\Lambda$CDM prediction \citep{bull01a}.  However,
if we attempt to add a component to the mass profile representing the
stellar mass of the cD galaxy with a reasonable stellar mass-to-light
ratio, the fit is degraded substantially in the central $\sim
50$~kpc. Modifying the dark matter halo as the result of adiabatic
contraction arising from the early condensation of stellar baryons in
the cD \citep[e.g.,][]{blum86a,gned04a} further degrades the fit.

If instead we use the Sersic-like profile proposed by \citet{nava04a}
to represent CDM halos, then sizable stellar mass-to-light ratios are
implied that are reasonably consistent with predictions from single
burst stellar population models. However, the inverse Sersic index,
$\alpha\sim0.5$, obtained from the fits is a factor of $\sim3$ higher
than predicted; the dark matter profile in the core is inferred to be
shallower than CDM. Hence, we are unable to obtain a good fit with a
model consisting of CDM halo and a separate contribution from the
stars in the cD galaxy with a reasonable stellar mass-to-light ratio.

The good fit of the NFW profile to the total gravitating matter of
regular galaxy clusters appears to be a common feature of X-ray
studies. The bright, highly relaxed cluster A2029 follows the NFW
profile all the way down to $\approx 0.001\rvir$ \citep{lewi03a}.
Dedicated studies of small samples of other mostly relaxed clusters
with \chandra\ and \xmm\ \citep{poin05a,vikhlinin06} also do not report
significant deviations from the NFW profile arising from central
stellar mass, except for some lower mass, group-scale objects
\citep{vikhlinin06,gastaldello}.

Since it seems unlikely that deviations from hydrostatic equilibrium
in the hot ICM have conspired in the same way in all of these
(regular) clusters to produce a mass profile consistent with NFW in
the center, we conclude that the adiabatic contraction scenario does
not appear to describe the formation of X-ray clusters. In particular,
for A2589 we have estimated the amount of non-thermal pressure support
in the hot ICM from random turbulent motions and magnetic fields and
conclude that neither source, especially turbulence, can reconcile the
\xmm\ mass profile with the adiabatic contraction scenario assuming a
reasonable stellar mass-to-light ratio in the cD. We suggest that
X-ray observations of A2589 and other relaxed clusters favor the
scenario where processes during halo formation, such as the heating of
the dark matter by dynamical friction with member galaxies,
counteracts adiabatic compression and leads to a total gravitating
mass profile consistent with the pure NFW profile (e.g.,
\citealt{elza04a}; see also \citealt{loeb03a}).

\acknowledgments

We would like to thank Oleg Gnedin for kindly providing us his
adiabatic compression code. This research has made use of the
NASA/IPAC Extragalactic Database (NED) which is operated by the Jet
Propulsion Laboratory, California Institute of Technology, under
contract with the National Aeronautics and Space Administration. We
are grateful to acknowledge partial support from NASA grant
NNG04GL06G.

\bibliography{a2589_7,dabrefs}
\end{document}